\documentclass[12pt,showpacs,amsmath,amssymb,aps,floats,aps,floatfix,nofootinbib,12pt]{revtex4}%
\usepackage{epsfig}
\usepackage{dcolumn}
\usepackage{bm}
\usepackage{amsmath}
\usepackage{amsfonts}
\usepackage{amssymb}
\usepackage{graphicx}
\usepackage{color}
\usepackage{latexsym}
\def\nnd{\end{document}}

\def\be{\begin{equation}}
\def\ee{\end{equation}}

\newcommand{\bea}{\begin{eqnarray}}
\newcommand{\eea}{\end{eqnarray}}
\newcommand{\bwt}{\begin{widetext}}
\newcommand{\ewt}{\end{widetext}}

\def\u
\def\hZ{\widehat Z}

\def\eed{\end{document}}

\def\m_z{m_{\textrm {Z}}}

\renewcommand{\u}{\rm{u}}

\def\be{\beta}

\def\sl#1{#1\!\!\!\!/}
\def\rm#1{\textrm{#1}}

\makeatother
\begin{document}
\title{Probing Light Stop Pairs at the LHC}
\author{Xiao-Jun Bi$^1$}
\author{Qi-Shu Yan$^2$}
\author{Peng-Fei Yin$^1$}
\affiliation{$^1$Laboratory of Particle Astrophysics, Institute of High Energy Physics, Chinese Academy of Sciences,
Beijing 100049, China}

\affiliation{$^2$College of Physics Sciences, Graduate University of Chinese Academy of Sciences, Beijing 100049, China}
\begin{abstract}
In this work, we study the light stop pair signals at the LHC. We explore the SUSY parameter space with non-universal gaugino and third generation masses at the GUT scale. Recent LHC SUSY search results based on 35pb$^{-1}$ and 1fb$^{-1}$ of data are implemented to put the limits on stop pair events. The dark matter relic density and direct detection constraints are also taken into account. Detailed simulations on the signals and background for some benchmark points are performed, and it is found that the stop pair signals usually escape the LHC search if the present cut conditions are used. We also explore the potential and sensitivity of ILC to probe such scenarios. It is found that the ILC can detect them with an integrated luminosity of a few tens of fb$^{-1}$.
\end{abstract}

\maketitle

\section{Introduction}
Supersymmetry (SUSY) is one of the most attractive extensions
beyond the standard model (SM) which offers a solution to the
hierarchy problem and grand unification in gauge interactions.
Moreover, the R-parity conserved SUSY models naturally provide a
lightest supersymmetric particle (LSP) which is neutral and stable
and can be a dark matter candidate badly needed in order to
interpret the cosmological observations. To guarantee the
electroweak symmetry breaking naturally, the size of SUSY soft
breaking terms is believed to be of the electroweak scale, of
which the light Higgs boson mass is typically less than about
$140$ GeV and the masses of sparticles are around a few hundred
GeV which are within the reach of LHC. The running of LHC can test
supersymmetry by discovering a light Higgs boson and SUSY
particles.

Search for supersymmetry is one of the prime targets of LHC.
Certainly the strong interaction sector of the SUSY models is the
most important place to discover SUSY due to the large cross
sections for the production of colored particles, such as squarks
and gluinos, at the hadron colliders. As one of typically
signature of such processes, the LSPs appearing in the final
states lead to a large missing energy (experimentally a large
missing transverse momentum). Therefore SUSY signature should show
up best in the jets plus large missing energy channel. Currently,
the most stringent bound is derived from this channel, which is
the dominant signature for gluino and squark production. With
1fb$^{-1}$ of data at the LHC, no SUSY signatures are detected. The
bound on the gluino and squark masses are set to be larger than
about 800 GeV for the mSUGRA model
\cite{Chatrchyan:2011zy,Aad:2011ib}. The recent release of LHC
measurements motivated quite a few new works on SUSY \cite{SUSY
PHENO}, where the implications of CMS and ATLAS results to various
SUSY models are discussed.

If nature chooses SUSY as its working principle, the null results
in the supersymmetric searches at hadronic colliders (both
Tevatron and LHC) may indicate that SUSY may hide from our probe
in some ways. For example, one way is that the SUSY might be
splitting in its color sector, i.e. color sparticles are very
heavy (say more than a few TeV) and their cross sections are
highly suppressed by masses. Then SUSY can only show up at LHC via
processes from its electroweak sector. In order to find SUSY, we
have to accumulate more data. In some worse cases, SUSY may even
be missed at LHC. For example, almost degeneracy of the masses of
chargino and neutralino leads to soft leptons which can escape the
detection at LHC \cite{Baer:2011ec}. If it is such a case, a
linear electron-positron collider or a muon collider should be
necessitated in order to unravel such a scenario.

Another possible way is that SUSY signature is buried deeply in the background events and can not been selected out with the current search approach. For example, in the compressed SUSY models \cite{Martin:2007gf,LeCompte:2011cn}, where the mass difference between the squark and LSP is small, only few soft jets can show up in the final states. Thus the signature is so similar to the background events that the currently SUSY search selection conditions can not separate the signal events.

The scenarios in which gluino and squarks of the third generation
are light but degenerate with LSP may also be hidden to the LHC
search
\cite{Ajaib:2010ne,Chen:2010kq,AdeelAjaib:2011ec}.
One of such a possibility is the light stop scenario
\cite{hep-ph/0007165,Balazs:2004bu,Carena:2005gc,Carena:2008mj,Carena:2008rt,Bornhauser:2010mw,Kats:2011it,Huitu:2011cp,Choudhury:2008gb,Johansen:2010ac,Kraml:2005kb,
Gogoladze:2011be,Essig:2011qg,Kats:2011qh,Brust:2011tb,Papucci:2011wy,arXiv:1111.2830,arXiv:1111.4467}.
The stop can be the next-to-lightest supersymmetry particle
(NLSP), which can be quite naturally realized in SUSY models, like
mSUGRA. The large top Yukawa coupling possesses two-fold effects
to the stop quark mass spectra. First, the off-diagonal trilinear
term, which is proportional to top Yukawa coupling, can lead to
the largest mass splitting in squark sector and consequently
produce a light stop as the NLSP. Second, since the lightest stop
is mostly right-handed, the mass of the right-handed stop,
$m_{\tilde t_R}$, is significantly reduced by the top Yukawa
coupling via the renormalization group running. Light stop scenario is also well-motivated to explain the dark matter relic density measured by WMAP when the mass splitting between it
and the dark matter candidate (the LSP) neutralino ($\chi_1^0$) is
small enough \cite{Carena:2005gc}.  Such a case is also dubbed as
the stop-neutralino coannihilation scenario. Furthermore, the
electoweak baryogensis in the framework of MSSM favors a light
stop \cite{Balazs:2004bu,Carena:2008rt}, i.e. the light stop can generate the
first order phase transition which prevents the baryon asymmetry
of the universe from being washed out. In order to guarantee this
first order phase transition, the mass of the light stop should be
light.

In the mSUGRA scenario, all the gaugino and scalar masses are
usually assumed universal at the GUT scale. However, such
an assumption is not guaranteed by symmetry. If the F-term of the gauge kinetic function in the SU(5) SUSY GUT
is not singlet, the gaugino masses at GUT would not be universal
\cite{Ellis,Drees}. The non-universal gaugino masses can also be predicted in the supersymmetric $SU(5)\times SU(3)_{Hypercolor}$ model proposed to solve Higgs doublet-triplet splitting problem in SU(5) GUT \cite{Yanagida:1994vq,ArkaniHamed:1996jq}, and in the supersymmetric partial unified model $SU(4)_C\times SU(2)_L\times SU(2)_R$\cite{shafi}. For the scalar sector, notice the most stringent constraints on SUSY FCNC and CP violation processes only relate to first two generations sfermions, it is well motivated to consider the so-called "inverted scalar mass hierarchy" scenario \cite{Dimopoulos:1995mi,Pomarol:1995xc,Cohen:1996vb,Barger:1999iv}. In such scenario, the large masses of first two generations sfermions can solve the SUSY flavor and CP problems, while the third generation sfermions are still light to satisfy the naturalness conditions. These non-universal soft breaking parameters
will change the running behavior of the RGE, and induce different
sparticle mass spectra and search strategies from the mSUGRA.

In this work, we will study a SUSY scenario with non-universal gaugino and third generation masses at the GUT scale, and then
explore the signatures of the light stop pair production (for
some relevant studies, see
\cite{Kawamura:1994ys,Nath,Chamoun:2001in,nonuniversal,Bhattacharya:2011bm,Feldman:2007zn}).
We first scan the SUSY parameter space in the non-universal SUSY
model. We have considered all the constraints on the SUSY parameter
space. Present 1 fb$^{-1}$ of LHC data at ATLAS
and CMS are used to put limits on stop pair events. We also simulate the signatures in some benchmark points at the LHC in detail. In general, the stop pair productions
depend on stop mass parameter, while the light stop decay modes only rely
on light slepton, chargino and neutralino mass spectra. Our results can be easily
extended to the scenarios with decoupled gluino and first two
generation squarks. For the stop-neutralino coannihilation
scenario to give correct dark matter relic density we find the
SUSY signatures are hidden by the present cut conditions. We
further present a study of the signals at the future linear
collider machine.

This paper is organized as follows. In Section 2 we start with the
generic bounds on SUSY from LEP, Tevatron, and LHC, while we
concentrate on the bounds to the light stop from its various decay
channels searched by experiments. In Section 3 we analyze the dark
matter bounds to the parameter space of non-universal SUSY models
and study the mass spectrum. In  Section 4 we discuss the recent
LHC bounds for our light stop scenario, and select four bench mark
points, demonstrate their mass pattern and decay features, and
elaborate how the signature from these bench mark points can hide
from the current search at LHC. We also study the sensitivity of
ILC to these bench mark points. Section 5 is for discussions and
conclusions.

\section{Current experimental bounds on light stop scenario}

In this section we describe the available bounds from the stop
pair search at LEP, Tevatron and LHC in details. Several stop
decay modes are investigated in \cite{Demina:1999ty}, i.e. 1)
${\tilde t} \to \chi_1^+ b$, 2) ${\tilde t} \to {\tilde \nu} \ell
b$, 3) ${\tilde t} \to {\tilde \ell^+} \nu b$, and 4) ${\tilde t}
\to \chi_1^0 c$. The recent bounds listed below depend on the mass
spectrum and the decay mode of the sparticles.

\begin{itemize}
\item ${\tilde t_1} \to b \chi^+ \to b \ell \nu \chi_1^0$: The pair production of stop decaying via ${\tilde t_1} \to b \chi^+$ has been investigated by the CDF collaboration with an integrated luminosity of 2.7 $fb^{-1}$ \cite{cdf2010}. The stop masses between 128 and 135 GeV are excluded at $95\%$ independent of the branching ratio of $\chi^+\to  \ell \nu \chi_1^0$. For $m({\tilde \chi_1^0})=45$GeV, $m(\chi_1^\pm)=125.8$GeV and $Br(\chi^+\to  \ell \nu \chi_1^0)=1$, lower limit for stop mass can be set at 196 GeV.
\item ${\tilde t_1} \to b \ell {\tilde \nu}$: The most stringent bounds for stop pair production with this decay channel are given by the D0 collaboration \cite{d02011} with an integrated luminosity of 5.4 $fb^{-1}$. The main final states are focused on $b\bar{b} e^\pm \mu^\pm {\tilde \nu} {\bar{\tilde \nu}}$. The sneutrino is assumed to be the LSP or decay invisibly into $\nu {\tilde \chi_1^0} $. The analysis is optimized for the mass region $\Delta m = m_{{\tilde t}_1}-m_{\tilde \nu} = 40 GeV$ or above. Stop masses up to 240 GeV are excluded for sneutrino masses around to 45 GeV, and sneutrino masses up to 120 GeV are excluded for stop masses around to 180 GeV.
\item ${\tilde t_1} \to t \chi_1^0$: The recent search for the top partner $T^\prime$ (it can be the $t^\prime$ of the fourth generation model, the new heavy quark in the little Higgs model with T-parity, or the scalar top quark in the SUSY model) and $T^\prime \to t \chi$ via the semi-leptonic mode $pp\to \ell \nu_{\ell} b q q^\prime b + \chi \chi$ and full-hadronic final states $pp\to q_1 q_2 b q_3 q_4 b + \chi \chi$ by CDF collaboration are reported from the Ref. \cite{Aaltonen:2011rr} and \cite{Aaltonen:2011na}, with integrated luminosities of 4.8 $fb^{-1}$ and 5.7 $fb^{-1}$ respectively. It is shown that the mass of $T^\prime$ can be bound up to 360 GeV for $m_X < 100 \textrm {GeV}$ and 400 GeV for $m_X \leq 70 \textrm {GeV}$, respectively. When both top quarks decay semi-leptonically, the final states are the same as the first two decay modes, i.e. ${\tilde t_1} \to b \chi^+ $ and ${\tilde t_1} \to b \ell {\tilde \nu}$. The discovery of this decay channel with both the discovery hadronic and semi-leptonic channels can distinguish this decay mode from the others.
\item ${\tilde t_1} \to c \chi_1^0$: If the the above processes are all kinetically forbidden, the loop induced flavor changing process ${\tilde t_1} \to c \chi_1^0$ might be dominant decay channel. The similar process ${\tilde t_1} \to u \chi_1^0$ is always suppressed by the CKM matrix when ${\tilde t_1} \to c \chi_1^0$ kinematically opens. This channel is difficult to be detected if the mass splitting between ${\tilde t_1}$ and $\chi_1^0$ is smaller which leads to two soft jets in the final states. The CDF collaboration had performed the research for this process with 2.6 $fb^{-1}$ of data \cite{CDFtjchi}. At least one of the jets is required to be tagged from a heavy-flavor quark. The analysis is optimized for the mass region $\Delta m = m_{{\tilde t}_1}-m_{\tilde \nu} = 40 GeV$ or above. Stop masses up to 180 GeV are excluded for neutralino masses around to 95 GeV. In addition, the results from LEP had excluded stop masses up to 90GeV for ${\tilde t_1} \to c \chi_1^0=1$. A more recent experimental search at D0 collaboration can be found in \cite{Shamim:2008zz}.
\item ${\tilde t_1} \to b f f^\prime \chi_1^0$: For small mass splitting between stop and the LSP, the four body process ${\tilde t_1} \to b f f^\prime \chi_1^0$ could be also important \cite{Boehm:1999tr}. Typically, the objects in this channel are softer comparing with those from ${\tilde t_1} \to t \chi_1^0$.
\item R-hadron : If the life-time of stop is long-enough due to the extremely small mass splitting or weak interaction between stop and the LSP \cite{Choudhury:2008gb}, stop can form a bound state R-hadron in the process of hadronization before its decay. If such R-hadron carrying electric charge, it might be observed in the inner tracker and even outer muon detector. Recently, CMS collaboration reported the limits for R-hadrons from pair production of stable stops based on 1.09 $fb^{-1}$ of data \cite{CMSRH}. Lower limit for stop mass can be set at 620GeV and 515GeV, corresponding to whether R-hadrons can leave observable signatures at the muon detector or not respectively. The search strategy of long-lived stops in MSSM is studied at LHC and can be found in the Monte Carlo study by Ref. \cite{Johansen:2010ac}.

\end{itemize}

\section{Analysis of the parameter space}

\subsection{Theoretical scenarios}

In the mSUGRA scenario, all the gaugino masses are set to be universal as $m_{1/2}$ at the GUT scale. However, this is a convenient assumption rather than a theoretical requirement. A non-universal gaugino mass sector is well motivated in many superstring or SUSY GUT models \cite{Ellis,Drees,Kawamura:1994ys,Nath,Chamoun:2001in}. In the SUSY GUT, if there exists a holomorphic function $f(\Phi)$ in the gauge kinetic function, gauginos will acquire masses via non-zero F-term of the $f(\Phi)$
\begin{equation}
\int d^2 \theta \; f(\Phi)_{ab} W^a W^b + h.c. \;\; \sim \;\;  \frac{1}{\Lambda} \langle F(\Phi)\rangle _{ab} \lambda^a \lambda^b .
\end{equation}
where $\Phi$ is a chiral field in the hidden sector related to
SUSY breaking, $\langle F(\Phi) \rangle_{ab}$ transforms in the
$Adj\otimes Adj $ representation of the underlying gauge group
containing gauginos $\lambda^a $, $\lambda^b$. For SU(5) SUSY GUT,
$\langle F(\Phi) \rangle_{ab}$  belongs to  $24 \otimes 24 = 1
\oplus 24 \oplus 75 \oplus 200$. Only if $\langle F(\Phi)
\rangle_{ab}$ is a singlet as $\langle F(\Phi) \rangle_{ab} \sim c
\delta_{ab}$, gaugino masses are universal as in the mSUGRA. The
non-universalities are generated when $\langle F(\Phi)
\rangle_{ab}$ belongs to other high representations. For example,
the gaugino mass relationships at GUT scale are $2:-3:-1$,
$1:3:-5$ and $1:2:10$ for $f(\Phi)$ belonging to $24$, $75$, $200$
respectively \cite{Ellis,Drees}. Moreover, if different $f(\Phi)$
representations appear simultaneously, other gaugino mass
relations can be achieved.

In the scalar sector, a general K$\ddot{a}$hler potential could also leads to mass non-universality \cite{Kawamura:1994ys,Nath}
\begin{equation}
\int d^2 \theta d^2 \bar{\theta} \; K_0 Q^{\dagger} Q + K^i_j Q^{\dagger}_i Q^j + ... \;\; \sim \;\; c_{ij} \phi^i \phi^j\,,
\end{equation}
where $K_0$, $K^i_j$, ... are real function of $\Phi^\dagger$, $\Phi$. In the mSUGRA scenario, all the scalar fields acquire same mass under the universal K$\ddot{a}$hler potential assumption $K^{i}_j \sim K \delta^i_j$. This assumption is useful to suppress FCNC effects which are stringently constrained by experiments. However, solving the SUSY flavor and CP problems only require large masses of first two generations sfermions, while the third generation sfermions can still be light to satisfy the naturalness conditions \cite{Dimopoulos:1995mi,Pomarol:1995xc,Cohen:1996vb,Barger:1999iv}.

Therefore, in this work, we treat these soft breaking mass parameters as free parameters at the GUT scale without imposing specific relations among them which can be derived from their underlying non-universal models. For the sake of simplicity, we choose the following seven input parameters as free parameters in our analysis:
\begin{equation}
M_{1/2}, \; M_{1/2,3},\; m_{0},\; m_{0,3},\; A_0,\; \tan \beta,\;
sign (\mu)\,.
\end{equation}
Compared with mSUGRA, only two extra input parameters at GUT
scale, i.e. the gluino mass $M_{1/2,3}$ and the third generation
sfermion mass $m_{0,3}$, are added.

\subsection{Parameter Space Scan and Experimental Constraints}

In this subsection we scan the SUSY parameter space with the 7
free parameters at the GUT scale discussed above. The low energy
spectra are calculated by solving the RGEs. In the scanning the
GUT and the electroweak symmetry breaking scale are set to be
$M_{GUT}=2 \times 10^{16}$ GeV and $M_{EWSB}=\sqrt{m_{\tilde{t}_1}
m_{\tilde{t}_2}}$, respectively. We consider the case where gluino
is lighter than wino and bino at $M_{GUT}$ and put upon a
constraint by requiring $100 \textrm{GeV} <M_{1/2,3}< M_{1/2}< 800
\textrm{GeV}$. To obtain a lighter stop or stau which is necessary
in order to yield a suitable DM relic density, we further assume
that third generation sfermions are lighter than the other scalars
in the first two generations, and allow them to vary in the range
$100 \textrm{GeV} <m_{0,3}<m_0< 2000
\textrm{GeV}$. The trilinear
coupling $A_0$ and the ratio of vacuum expectation values
$\tan\beta$ are chosen in the range of $ -1<A_0/m_0<1$ and
$2<\tan\beta<50$, respectively. The sign of $\mu$ is taken to be
positive, which is favored by several experimental constraints on
$b\to s \gamma$ etc. We utilize \textsf{SuSpect} \cite{suspect} to
calculate the SUSY particles spectra by solving the two-loop SUSY
renormalization group equations. The top quark pole mass can
affect the sparticle spectra and consequently modify dark matter
relic density significantly. We take the top pole mass as
$m_t(\text{pole})=173.1$ GeV.

Several phenomenology and astrophysics experimental constrains are taken into account in our scanning. Important flavor physics constraints: $Br(b\to s\gamma)=(3.55\pm0.24)\times 10^{-4}$ \cite{bsgamma}, $Br(B_s\to \mu^+ \mu^-)=(0\pm1.4)\times 10^{-8}$ \cite{bsmumu}, $Br(B_u\to \tau \bar{\nu})/SM=1.28\pm0.38$ \cite{bsgamma}, are realized. As a conservative analysis, we only demand that the SUSY contributions pass these constraints to a $3\sigma$ level. Another remarkably important constraint is from the muon anomalous magnetic moment $g_\mu-2$ measurement \cite{gmi2}, which is adopted here as a bound $-11.4\times 10^{-10}< g_\mu-2 < 9.4 \times 10^{-9}$ as used in Ref. \cite{Feldman:2007zn}. The lighter Higgs boson is required to be heavier than 114 GeV, while the mass limits for other charged sparticle from LEP are also imposed \cite{PDG10}.

Constraints by dark matter relic density is also considered. In
our analysis, the lightest neutralino is required to be LSP and a
candidate of dark matter. The dark matter relic density is
reported by WMAP7 in the range $\Omega h^2= 0.112\pm0.0056$
\cite{Larson:2010gs}. However, we only require the thermal
abundance of neutralino satisfies a $3\sigma$ upper-bound
$\Omega_{\chi^0_1} h^2 < 0.1288$ due to the reason that the
neutralino might not be the only dark matter particle or that the
neutralino might be produced in early universe via the so-called
non-thermal mechanism \cite{xmzhang2000}. The recent direct search
experiment XENON100 \cite{{Aprile:2011hi}} is realized, which put
the most stringent bound upon the dark matter-nucleon
spin-independent scattering. We demand that the reduced dark
matter-nucleon scattering cross section
$\sigma^{SI}_r=\sigma^{SI}_{\chi^0_1 p} (\Omega_{\chi^0_1} h^2 /
\Omega h^2)$ should be smaller than XENON100 limit. All
constraints on flavor physics from low energy colliders and all
bounds on dark matter from astrophysics and direct detection are
implemented by using \textsf{MicrOMEGAs} \cite{micromega} which
uses the \textsf{SuSpect} output as input parameters.

\subsection{Numerical results and sparticle masses}

In this section we present some results based on a $\sim 10^6$ case study in our parameter space scan. We find that most of the points can not yield correct RGE solutions nor induce the spontaneous electroweak symmetry breaking. Only $\sim 700$ points can pass all the constraints. To analyze the features of dark matter and collider signatures, it is found that light sparticles play the crucial roles. It is convenient to categorize the parameter points in term of the mass hierarchical relation of light sparticles as suggested in Ref. \cite{Feldman:2007zn}. Here we define our four mass patterns:

(1) the stop pattern (SO): $m_{\chi^0_1}< m_{\tilde{t}_1}< m_{\tilde{\tau}_1}, m_{\chi^\pm_1}$ ;

(2) the stau/stop pattern (SS): $m_{\chi^0_1}< m_{\tilde{\tau}_1}< m_{\tilde{t}_1}< m_{\chi^\pm_1}$ ;

(3) the stau pattern (SA): $m_{\chi^0_1}< m_{\tilde{\tau}_1}< m_{\chi^\pm_1}< m_{\tilde{t}_1}$ ;

(4) the chargino pattern (CH): $m_{\chi^0_1}< m_{\chi^\pm_1}< m_{\tilde{\tau}_1}, m_{\tilde{t}_1}$.

\begin{figure}[!htb]
\begin{center}
\includegraphics[width=0.47\columnwidth]{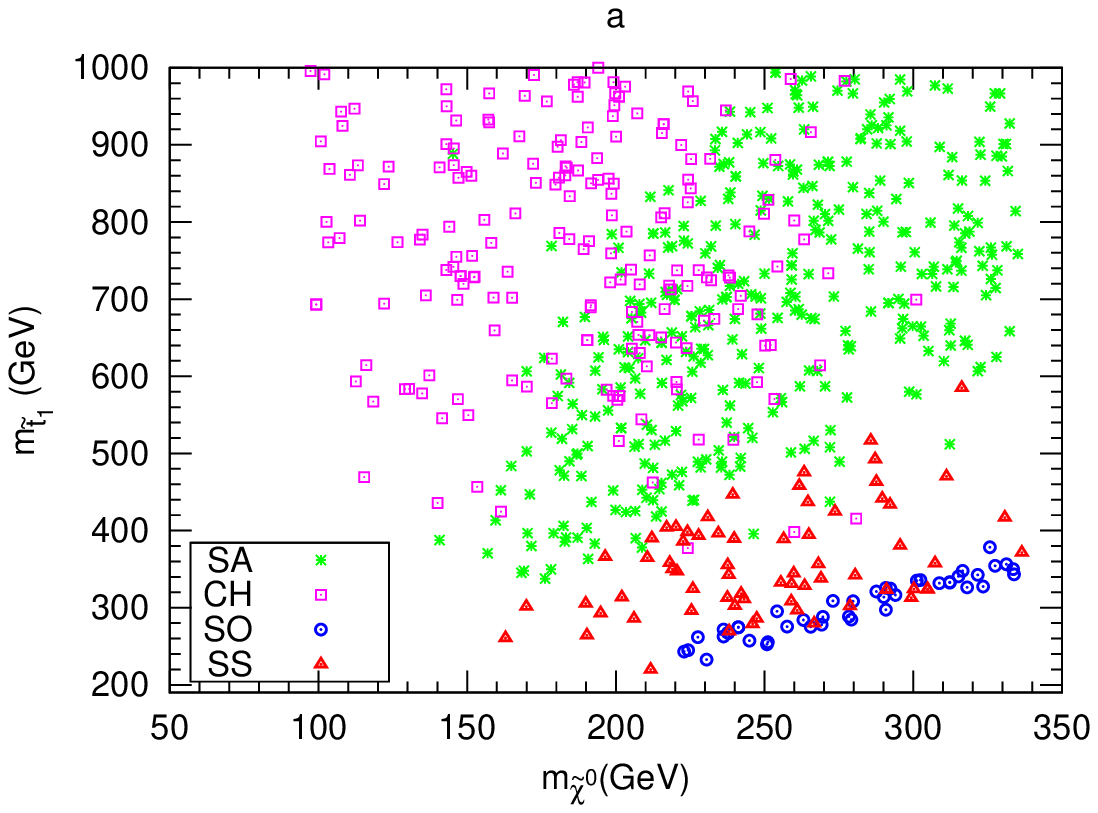}
\includegraphics[width=0.47\columnwidth]{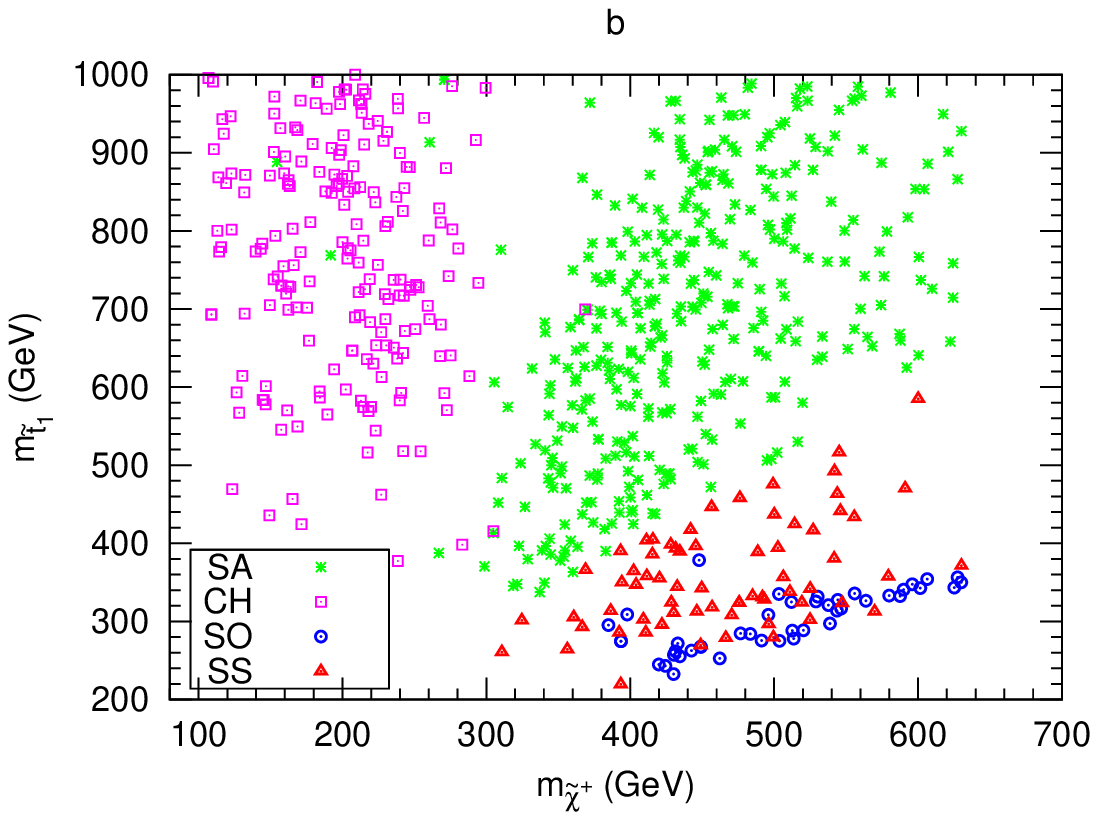}
\caption{Four mass patterns in the planes (a) $m_{\chi^0_1}$ vs
$m_{\tilde{t}_1}$ (top left), (b)  $m_{\chi^\pm_1}$ vs
$m_{\tilde{t}_1}$ (top right) are illustrated.
\label{mass}}
\end{center}
\end{figure}

\begin{figure}[!htb]
\begin{center}
\includegraphics[width=0.47\columnwidth]{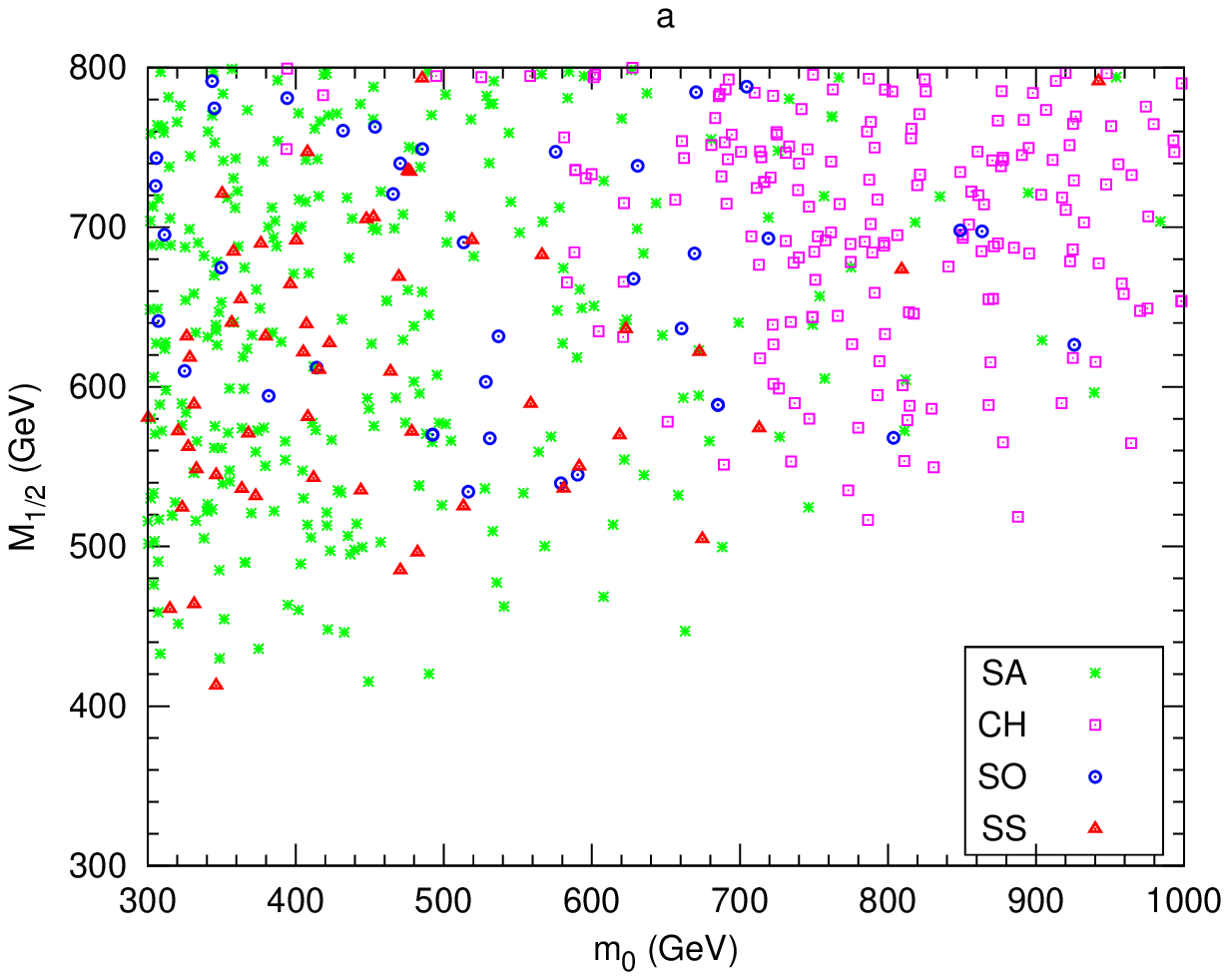}
\includegraphics[width=0.47\columnwidth]{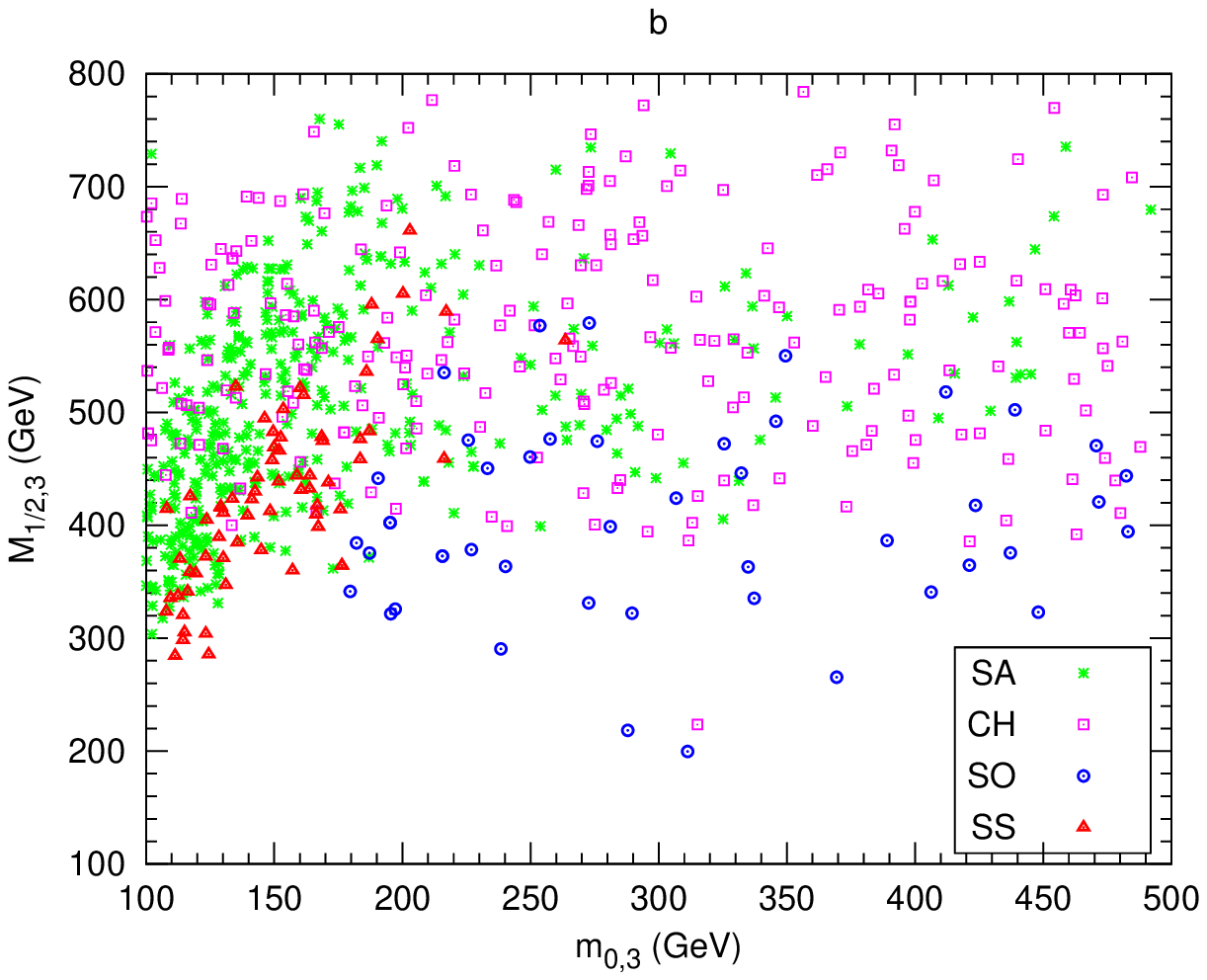}
\includegraphics[width=0.47\columnwidth]{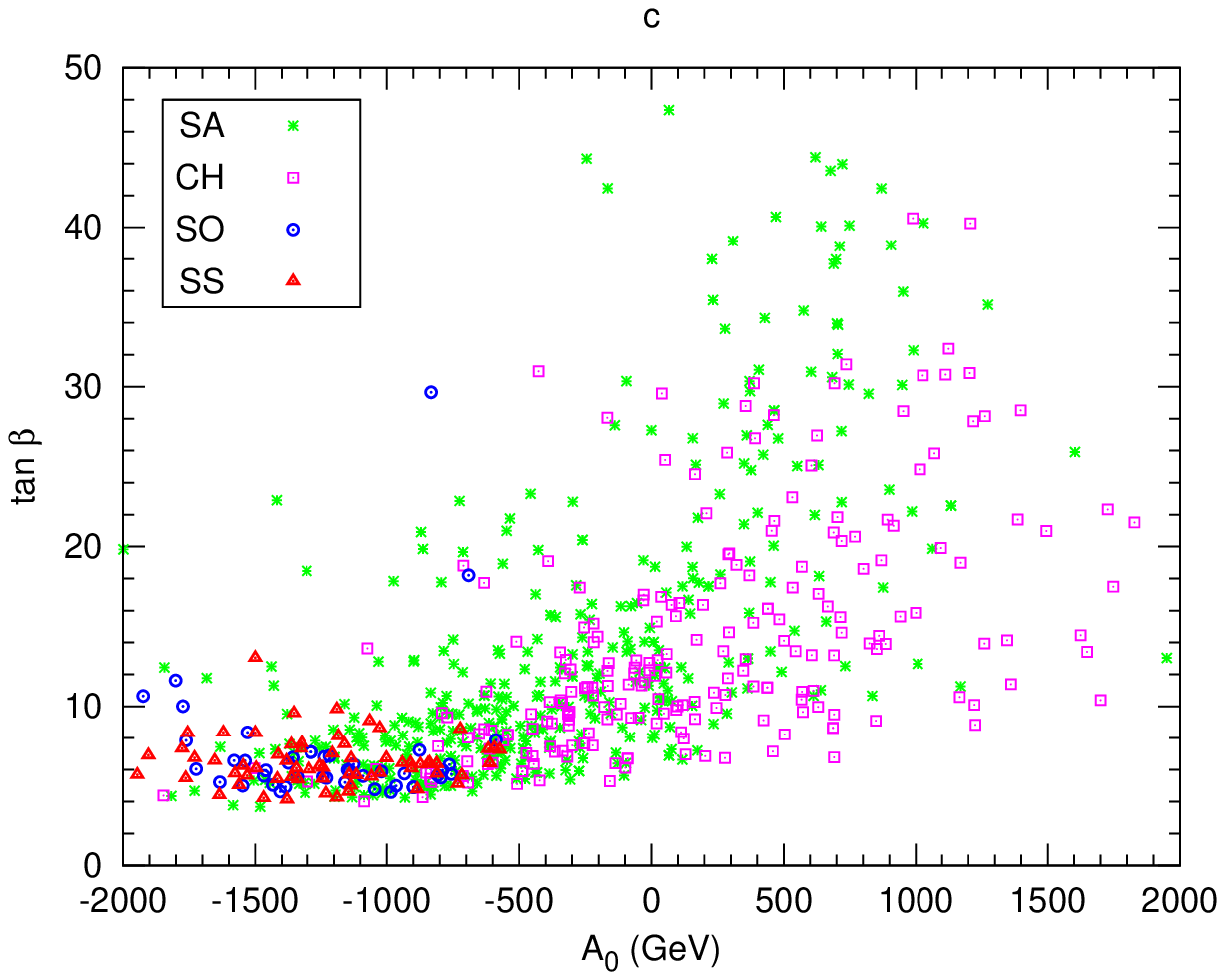}
\includegraphics[width=0.47\columnwidth]{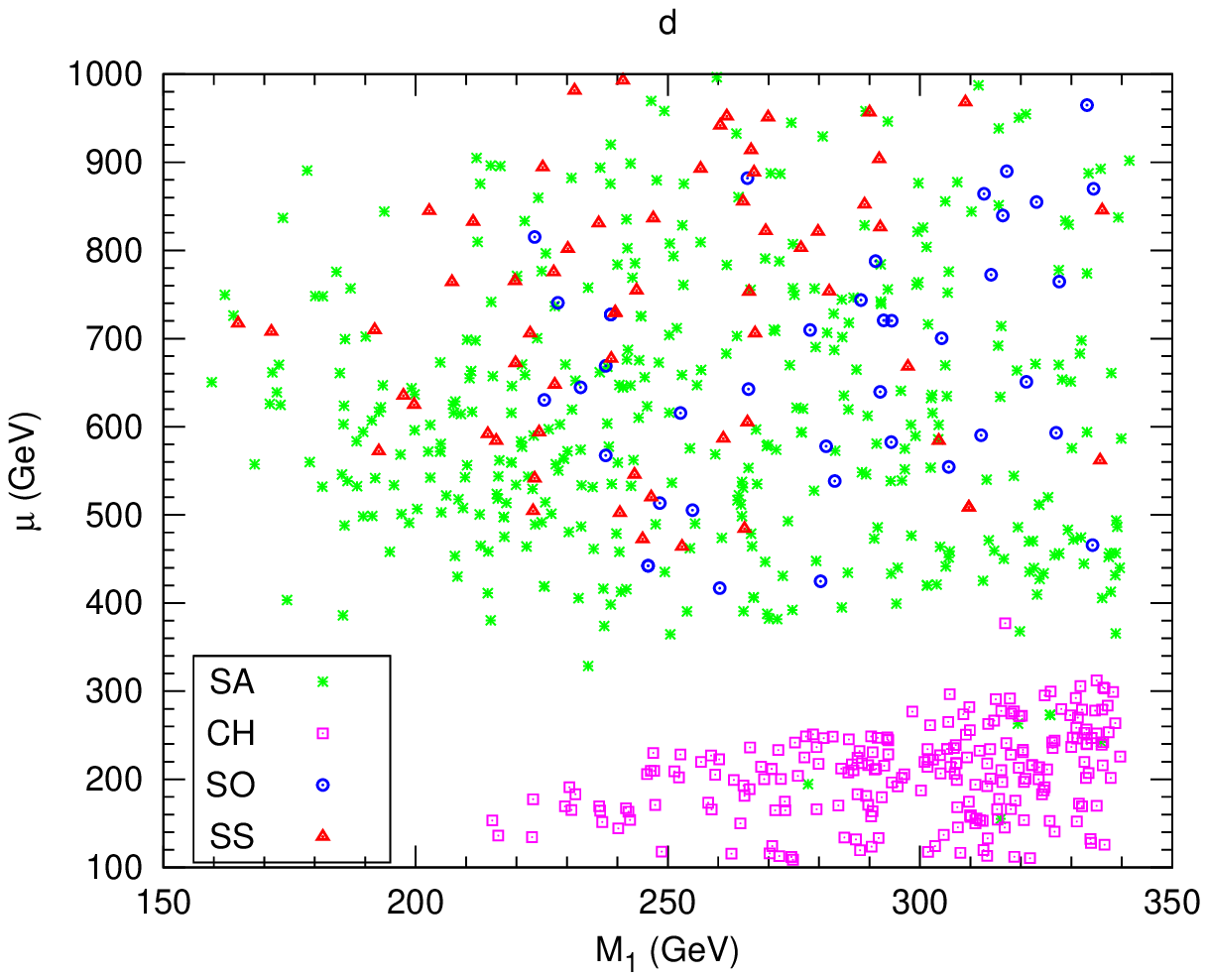}
\caption{Four mass patterns in the planes (a) $m_0$ vs $M_{1/2}$ (top left), (b)  $m_{0,3}$ vs $M_{1/2,3}$ (top right), (c) $A_0$ vs $\tan \beta$ (bottom left), (d) $M_1$ vs $\mu$ (bottom right) are displayed, respectively.
\label{soft}}
\end{center}
\end{figure}

In Fig. \ref{mass}, we present two scattering plots to reveal some
features of these four mass patterns. In the plot (a), points are
shown in the $m_{\chi^0_1}- m_{\tilde{t}_1}$ plane. In the plot
(b), they are shown in $m_{\chi^\pm_1}-m_{\tilde{t}_1}$ plane. The
distribution of these four mass patterns in soft SUSY breaking
parameter space are also shown in Fig. \ref{soft}. In Fig.
\ref{soft}(a), Fig. \ref{soft}(b), Fig. \ref{soft}(c), and Fig.
\ref{soft}(d), the distribution on the $m_0-M_{1/2}$ plane, on the
$m_{0,3}-M_{1/2,3}$ plane, on the $A_0-\tan \beta$ plane, and on
the $M_1-\mu$ plane are displayed, respectively. Four comments on
the distributions are listed in order:

(1) The gaugino mass running behavior in the non-universal scenarios is similar to that in mSUGRA. The 1-loop RGEs for gaugino masses are
\begin{equation}
16 \pi^2 \frac{d}{dt} M_i=2 b_i g_i^2 M_i\,\, ,
\end{equation}
where $(b_1,b_2,b_3)=(33/5,1,-3)$. It is well-known that in the mSUGRA the soft breaking parameters $M_3^0=M_2^0=M_1^0$ at $M_{GUT}$ evolves to $ M_3:M_2:M_1 \sim \alpha_3 M_3^0:\alpha_2 M_2^0:\alpha_1 M_1^0 \sim 6M_3^0: 2M_2^0: M_1^0$ at $M_Z$. In the non-universal scenarios, gaugino mass parameters of first two generations at low energy hold this relation $M_1:M_2 \sim 1:2$, which consequently means that the main component of the lightest neutralino $\tilde{\chi}^0_1$ is either Bino or Higgsino, depending on the relations of $M_1$ and $\mu$. This $M_1-M_2$ relation also implies the lighter chargino $\tilde{\chi}^+_1$ and the next lightest neutralino $\tilde{\chi}^0_2$ should almost degenerate, since both of them are SU(2) gaugino dominant $m_{\tilde{\chi}^+_1} \sim m_{\tilde{\chi}^0_2} \sim M_2$. In the Fig. \ref{mass}(a) and Fig. \ref{mass}(b), we can observe that the $ m_{\tilde{\chi}^0_1}$ and $m_{\tilde{\chi}^+_1}$ vary in the region of $\sim$ (100, 350)GeV and (100, 700)GeV respectively, which basically reflects such a relation. On the other hand, the gluino mass parameter $M_3$ gains a large contribution via a negative $\beta$ function at low energy, which leads to large gluino mass. It is beyond the reach of Tevatron, or even the reach of LHC with $\sqrt{s}=7$TeV if gluino mass is heavier than $1$ TeV.

(2)The 1-loop RGEs for the third generation right-handed sfermion squared-mass parameter could be written as
\begin{eqnarray}
16 \pi^2 \frac{d}{dt} m_{\bar{u}_3}^2&=&4y_t^2(m_{H_u}^2+m_{Q_3}^2+m_{\bar{u}_3}^2)-\frac{32}{3}g_3^2 |M_3|^2-\frac{32}{15}g_1^2 |M_1|^2 +4|A_t Y_t|^2- \frac{4}{5}g_1^2 S , \\
16 \pi^2 \frac{d}{dt} m_{\bar{d}_3}^2&=&4y_b^2(m_{H_d}^2+m_{Q_3}^2+m_{\bar{d}_3}^2)-\frac{32}{3}g_3^2 |M_3|^2 -\frac{8}{15}g_1^2 |M_1|^2 +4|A_b  Y_b|^2+ \frac{2}{5}g_1^2 S ,     \\
 16 \pi^2 \frac{d}{dt} m_{\bar{e}_3}^2&=&2y_{\tau}^2(m_{H_d}^2+m_{L_3}^2+m_{\bar{e}_3}^2)-\frac{24}{5}g_1^2 |M_1|^2 +2|A_\tau  Y_\tau|^2+ \frac{6}{5}g_1^2 S ,
\label{rgesfermion}
\end{eqnarray}
where $S$ is defined as $S= Tr[Y_i m_{\phi_i}^2]$. For the first two generation sfermions, the terms proportional to Yukawa couplings can be neglected due to the tiny values of their Yukawa couplings. In contrast, the sfermion mass terms of the third generation can get a large contribution from the Yukawa coupling terms. Therefore a light stop is often the lightest squark due to its large Yukawa couplings. When compared with mSUGRA, the assumption $m_{0,3}< m_0$ is can further decrease the contribution of gluino to the squark mass parameters. This can be read out from Fig. \ref{soft}(b) where typically our stop pattern corresponds a smaller value in $m_{0,3}$ parameter. The lightest squark is the lighter stop means that its cross section can be the largest at the hadronic colliders, Tevatron and LHC. The mass relation between stop and stau is difficult to be read out from RGE. Typically, the $m_{\bar{\tau}}^2$ gets less positive contributions from gauginos and the Yukawa coupling $Y_\tau$ is much smaller than $Y_t$ as well, which leads to quite a few mass patterns of light third generation sfermions.

(3) The 1-loop RGE for the Higgs mass parameters $H_u^2$ is given as
\begin{equation}
16 \pi^2 \frac{d}{dt} m_{H_u^2}^2 = 6y_t^2(m_{H_u}^2+m_{Q_3}^2+m_{\bar{u}_3}^2)-6g_2^2 |M_2|^2-\frac{6}{5}g_1^2 |M_1|^2 +6|a_t|^2+ \frac{3}{5}g_1^2\,\,.
\end{equation}
The $m_{H_u}^2$ evolves to be negative at the low scale due to the large terms proportional to $Y_t^2$ which is essential in order to induce a spontaneous electroweak symmetry breaking. In the non-universal scenario, a smaller third generation sfermion mass leads to a $|m_{H_u}^2|$ smaller than its counterpart in the mSUGRA. A small $|m_{H_u}^2|$ can lead to a small soft Higgs mass parameter $\mu$, due to the fact that $\mu$ is determined by the $m_{H_u}^2$ from the tree level relation
\begin{equation}
\mu^2=\frac{m_{H_u}^2 sin^2 \beta -m_{H_d}^2 cos^2 \beta}{cos 2\beta } -\frac{m_Z^2}{2}.
\end{equation}
When the magnitude of $\mu$ is comparable to that of $M_1$, as
demonstrated in the cases of CH patten, the lightest neutralino $\tilde{\chi}_1^0$ can have a
large component of Higgsino due to the large mixing.

(4) The squared-mass matrix for the stop quark in the weak interaction eigenstate basis $(\tilde{t}_L,\tilde{t}_R)$ is given by
\begin{eqnarray}
\left(\begin{array}{cc}
\; m_{Q_3}^2+m_t^2+\Delta_{\tilde{u}_L} \;\;\;\;\;\;\;\;\;\; v(a_t^*sin\beta-\mu y_t cos\beta)\\ v(a_tsin\beta-\mu^* y_t cos\beta) \;\;\;\;\;\; m_{\bar{u}_3}^2+m_t^2+\Delta_{\tilde{u}_R}
\end{array}\right),
\end{eqnarray}
where $\Delta_{\phi_i}=(T_{3\phi}-Q_{\phi}
sin^2\theta_W)cos(2\beta)m_Z^2$, and $a_t = A_t Y_t$.  The
off-diagonal elements can generate a large mass split between two
stop mass-eigenstates, which are labeled as
$(\tilde{t}_1,\tilde{t}_2)$. For the case $sign(\beta)>0$ chosen
in this study, a small $tg \beta$ and a large $-a_t$ can produce a
large mass splitting and leads to a relatively lighter stop
$\tilde t_1$, which is reflected in Fig. \ref{soft}(c). In
contrast, if the $a_t$ is positive and the $\mu$ is small,
$m_{\tilde{t}_1}$ might be not light when compared with
$m_{\tilde{\chi}_+}$, which leads to the cases of the CH patten.

\subsection{Features of the dark matter}

In SUSY models, most parameter space leads to a neutralino relic
density which is too large to overclose the Universe. Depending on
the mass spectra, four working processes can be introduced to
produce a small relic density of neutralino
\cite{Jungman:1995df}:(1) all the sfermions are light, neutralinos
annihilate via t-channel sfermions exchange; (2)
$\tilde{\chi}_1^0$ has significant component of Higgsino or wino,
main annihilating channel is to gauge bosons or Higgs; (3)
neutralinos scatter with sfermions with nearly mass degeneracy
which is so-called "co-annihilation"; (4) neutralinos annihilate
via s-channel Higgs resonance with $2m_{\tilde{\chi}_1^0}=m_{A^0}$, or $m_{h^0}$, $m_{H^0}$.

When the stop $\tilde{t}_1$ is light, two main processes
can lead to a small neutralino abundance. The first is the
neutralino-stop coannihilation process  $\tilde{\chi}_1^0
\tilde{t}_1 \to tg/h_0$, and the other is $\tilde{\chi}_1^0
\tilde{\chi}_1^0 \to t \bar{t}$ by the t-channel via exchanging
the light stop when the kinematic is allowed. When the lighter
stau $\tilde{\tau}_1$ is light, the neutralino-stau coannihilation
can occur and make dominant contribution. This is the cases of SA and SS. It is
worthy of mentioning that the neutralino-stop coannihilation can
be significant for the case of SS.

For the case of CH, the lightest neutralino $\tilde{\chi}_1^0$ has
a large component of Higgsino due to the large mixing deduced by
the small $\mu$ and the large $m_0$, as shown in the
Fig.\ref{soft}. Two dominant annihilation processes can occur. (1)
The pair of neutralinos can annihilation into gauge boson and
Higgs boson via the processes $\tilde{\chi}_1^0 \tilde{\chi}_1^0
\to W^+ W^-, ZZ, Zh_0$ due to large Higgsino component. (2) A
neutralino and a chargino can coannihilate into the particles of
the SM via the processes $\tilde{\chi}_1^0 \tilde{\chi}_1^{\pm}
\to W^{\pm} Z/\gamma, f\bar{f}'$, which can occur due to the small
chargino mass and a large Higgsino component in chargino
$\tilde{\chi}_1^{\pm}$.

The last but not the least, the sfermion-sfermion
self-annihilations  processes may also be important, which can be
attributed to the fact that a large sfermion-sfermion
self-annihilation can
increase the effective total dark matter cross section when the
mass splitting between sfermion and neutralino is small. Such a
case can happen as we can read out from the effective cross
section at the dark matter frozen-out epoch \cite{Griest:1990kh}

\begin{equation}
\sigma_{eff}=\Sigma_{ij;kl} \sigma_{ij;kl} r_i r_j,  \;\;\;\;\; r_i=\frac{n_{eq}^i}{n_{eq}}=\frac{g_i}{g_{tot}}(1+\Delta_i)^{3/2} exp(-\Delta_i m_{\tilde{\chi}_1^0}/T),
\end{equation}
where $\Delta_i$ is defined as $\Delta_i=(m_i-m_{{\chi}_1^0})/m_{\tilde{\chi}_1^0}$.

\begin{figure}[!htb]
\begin{center}
\includegraphics[width=0.47\columnwidth]{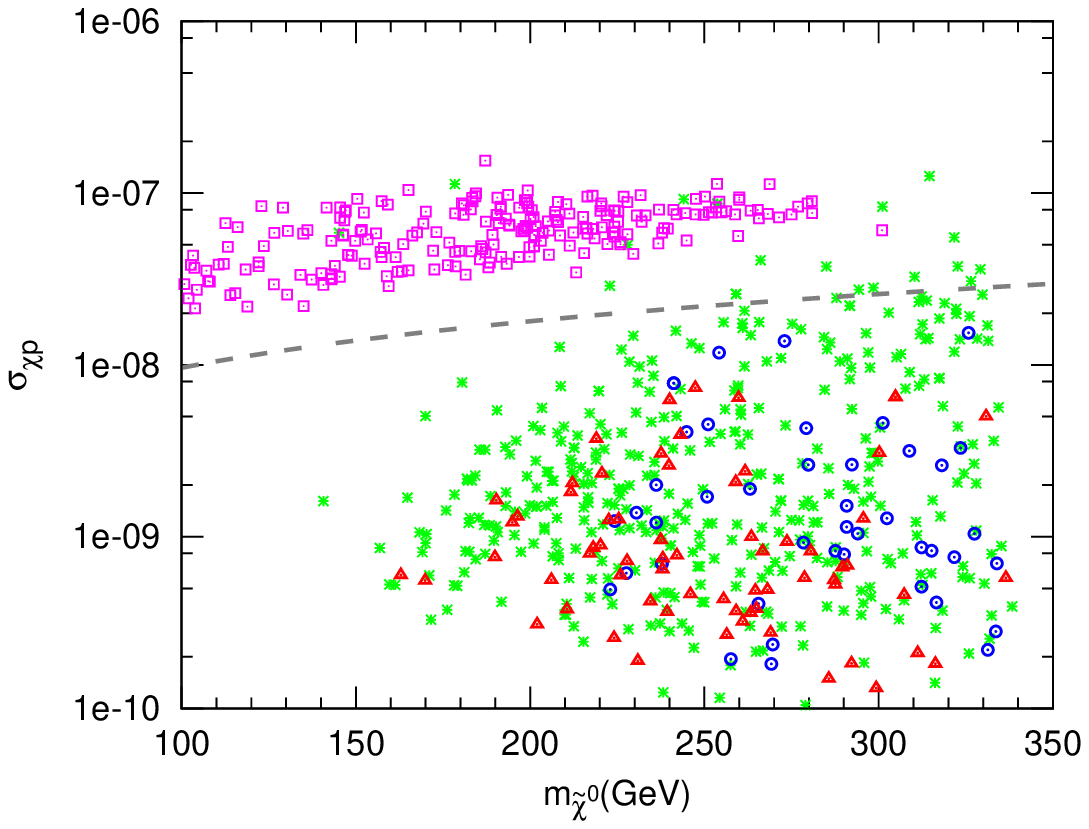}
\includegraphics[width=0.47\columnwidth]{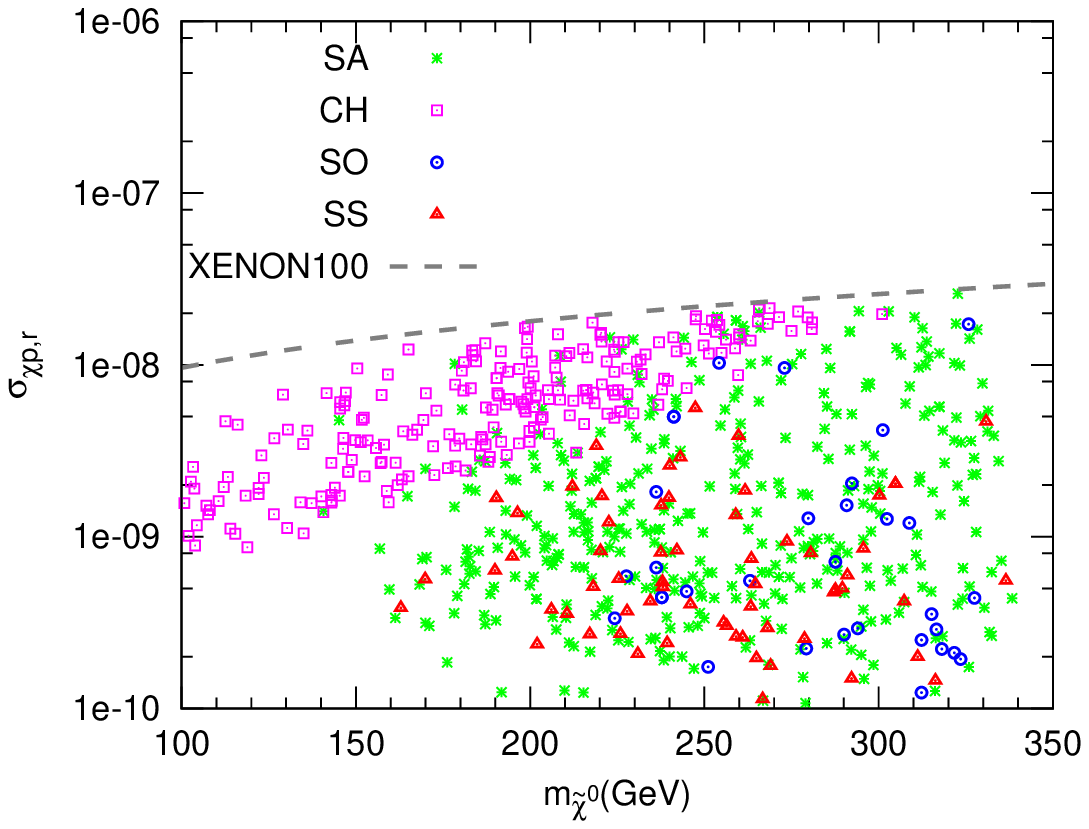}
\caption{The elastic spin independent DM-nucleon cross section $\sigma_{\tilde{\chi}_1^0p}$ (left panel) and reduced cross section $\sigma_{\tilde{\chi}_1^0p,r}$ (right panel) for our four mass patterns are
\label{ddx}}
\end{center}
\end{figure}

The neutralino-nucleon spin-independent scattering can be detected by dark matter direct search through $\tilde{\chi}_1^0 q \to \tilde{\chi}_1^0 q$ via squark or Higgs exchange. If the first two generation squarks are light or neutralino has a large Wino/Higgsino component, the null results from direct detection can put a strong constraint on SUSY models. Recently, the XENON100 collaboration observed 3 events with an background of $1.8\pm 0.6$ after 48kg$\times$100.9 days running \cite{Aprile:2011hi}. The result can be used to constrain the dark matter-nucleon spin-independent scattering cross section.

We show the neutralino-nucleon cross section $\sigma_{\tilde{\chi}_1^0 p}$ (left panel), the reduced cross section  $\sigma_{\tilde{\chi}_1^0 p, r}$ (right panel) and XENON100 limit in Fig.\ref{ddx}. Here the reduced neutralino-nucleon cross section defined as $\sigma_{\tilde{\chi}_1^0 p, r}=\sigma_{\tilde{\chi}^0_1 p} (\Omega_{\chi^0_1} h^2/\Omega h^2) $ takes into account that the neutralino may only contribute part of the total dark matter relic density. From Fig.\ref{ddx}(a), we can observe that the neutralino in the CH case has a large $\sigma_{\tilde{\chi}_1^0 p}$ due to its large Higgsino component. If we assume that the neutralinos are produced by a non-thermal process with a correct relic density $\Omega_{\chi^0_1} h^2 =\Omega h^2$, such a scenario is almost excluded by XENON100. If we assume the neutralino only contributes part of the total dark matter relic density, such a scenario is still allowed, as shown in Fig. \ref{ddx}(b).


\section{The signature of light stop pairs at the LHC}

\subsection{production and decay of stop pairs}

In this section, we focus on the production and decays of the
lighter top squark at the colliders. As pointed out above, the
lighter stop $\tilde{t}_1$ can have a smaller mass than all other
colored sparticles in the non-universal scenario and the mass
splitting between it and the other colored sparticle can reach to
$1$ TeV. Then it is expected that the cross section of the lighter
stop pairs at the LHC should be the largest one in our mass
patterns introduced above. A pair of stop can be produced via two
main processes via $q\bar{q}$ annihilation and gluon fusion
\cite{Beenakker:1997ut}. The first one is dominant at Tevatron
while the second is dominant at LHC.

\begin{figure}[!htb]
\begin{center}
\includegraphics[width=0.47\columnwidth]{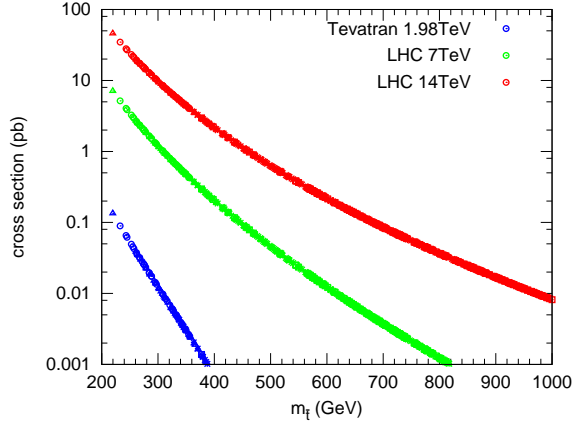}
\caption{The cross sections of stop pair production at Tevatron, LHC-7TeV and LHC-14TeV are shown.
\label{stoppro}}
\end{center}
\end{figure}

To evaluate the cross section of stop pair production $\sigma_{\tilde{t}_1 \tilde{t}_1}$ at hadronic colliders, we utilize the package \textsf{Prospino} \cite{Beenakker:1996ed} which has incorporated the next leading order corrections. The results for Tevatron with $\sqrt{s}=$1.98 TeV, LHC with $\sqrt{s}=$7TeV and 14TeV are plotted in Fig.\ref{stoppro}. The K factor is determined to be situated in the range of $\sim$(0.10, 0.13), $\sim$(0.15, 0.20) and $\sim$(0.15, 0.18) for Tevatron, LHC7 and LHC14, respectively.

In the Fig.\ref{stoppro}, we observe that the mass of
$\tilde{t}_1$ almost monotonically determines the cross section
$\sigma_{\tilde{t}_1 \tilde{t}_1}$, which is not sensitive to
other SUSY parameters. At Tevatron, it is 0.1pb or so for
$m_{\tilde{t}_1}\sim$ 200 GeV and quickly decreases to 1fb when
$m_{\tilde{t}_1}$ increases to 400 GeV. At the LHC with
$\sqrt{s}=$7TeV, the $\sigma_{\tilde{t}_1 \tilde{t}_1}$ is larger
than its value at Tevatron by a factor of 100  and decreases to
1fb for $m_{\tilde{t}_1}\sim$ 800GeV. For the
$m_{\tilde{t}_1}<$350GeV, more than one stop pair events at the
LHC7 with 35$pb^{-1}$ is predicted. With the increasing of
$\sqrt{s}$ to $14$ TeV and a larger integrated luminosity (say 100
fb$^{-1}$), stop pair events should be copiously produced when the
lighter stop is less than 400 GeV.

In order to analyze the signatures of stop pairs and to reduce the
standard model background by setting cuts, we have to study the
decay products of stop pair. For this purpose, we use
\textsf{SDECAY} package \cite{Muhlleitner:2003vg} to compute stop
decay branching fractions.

When the lighter stop is much heavier than the LSP, the two body decay $\tilde{t}_1 \to t \chi^0 $ and $\tilde{t}_1 \to b \chi^+_1 $ can be its dominant decay modes. In the $\tilde{t}_1 \to b \chi^+_1 $ decay mode, the $\chi^+_1 $ can  mainly decay into $\chi^0_1 W^+$ or $\nu \tilde{\tau}$, which is determined by the mass of stau and the dominant component of $\chi^+_1$. This is the feature of the SA and CH cases. When the lighter stop is smaller than $400$GeV, these two body decay modes may be forbidden kinematically. Then the three body decay modes $\tilde{t}_1 \to b W \chi^0_1 $ and $\tilde{t}_1 \to \nu \tilde{\tau} \chi^0_1$ become dominant \footnote{It is need to mention that the three body decay channels $\tilde{t}_1 \to l \tilde{\nu} \chi^0_1$ searched by Tevatron collaborations is not important in our scenario. Since the $m_{\tilde{\tau}}$ is much smaller than $m_{\tilde{\nu}}$, we observe that the branching fraction of $\tilde{t}_1 \to \nu \tilde{\tau} \chi^0_1$ is  always be larger than that of $\tilde{t}_1 \to l \tilde{\nu} \chi^0_1$} in our scan. When the mass splitting between stop and neutralino is too small to allow the three body decay, the four body decay mode $\tilde{t}_1 \to b jj \chi^0_1 / b \nu l \chi^0_1$ (a recent study showed that this decay mode can be used to probe trilinear coupling $A_0$ when tan$\beta$ is small \cite{Bhattacharyya:2011ew}) or the loop induced FCNC decay $\tilde{t}_1 \to c \chi^0_1$ become dominant.

\subsection{Simulation and analysis}

Recently, CMS and ATLAS collaborations reported several results on their searching for SUSY in different channels. In this section, we present the LHC bounds to the stop pair production with a collision energy $\sqrt{s}=7$TeV in term of different channels studied by CMS and ATLAS.

In our analysis, parton level events are generated by \textsf{MadGraph} \cite{Alwall:2007st}, while parton shower, decays, and hadronization are performed by \textsf{PYTHIA} \cite{Sjostrand:2006za}. \textsf{PGS} \cite{pgs} is used to simulate detector effects and to find jets, leptons, and missing transverse momentum. The acceptance cuts for all jets and charge leptons are chosen as $p_t> 20$GeV and $|\eta|<2.5$.

The SUSY search strategies of CMS and ATLAS are optimized for mSUGRA scenario at this stage. It always require large MET and energetic leading jets in order to capture the signature from the heavy squark or gluino pair productions. Therefore the search to the jets plus MET channel sets the most stringent constraint to ordinary parameter space in mSUGRA. Cuts for studying this channel are described as follows:

\begin{itemize}
\item In the Ref. \cite{Chatrchyan:2011zy}, CMS presents a search for SUSY signatures on an integrated luminosity of 1.14fb$^{-1}$ with jets and significant MET and without leptons. In this analysis, jets are required
\begin{equation}
E_T^{j_1,j_2}>100GeV,\; E_T^j>50GeV,\;\;\;|\eta|_j<3 ,
\end{equation}
where we have used the convention $p_T^{j_i}>p_T^{j_{i+1}}$. The following selections are adopted to compute event rate
\begin{equation}
H_T>275GeV,\;\;\; \alpha_T>0.55,\;\;\;
\end{equation}
where $H_T$ and $\alpha_T$ are defined as
\begin{equation}
H_T=\Sigma E_T^{j_i}, \;\;\; \sl{H}_T=|-\Sigma \vec{p}_T^{\;j_i}| \;\;\;
\label{HT}
\end{equation}
\begin{equation}
\alpha_T=E_T^{J_2}/\sqrt{H_T^2-\sl{H}_T^{\;2}}. \;\;\;
\end{equation}
If the jet number is more than 2, we utilize hemisphere algorithm \cite{cmstdr} to combine jets into two pseudo-jets named $J_1$, $J_2$ (assumed $p_T^{J_1}>p_T^{J_2}$ again). The events containing isolated leptons and photons are rejected, where isolated leptons and photons are objects with
\begin{equation}
p_T^\ell>10GeV,\;\; p_T^\gamma> 25GeV,\;\; |\eta|<2.5. \;\;\;
\end{equation}
CMS collaboration performed a search in eight bins of $H_T>275$GeV, and found the standard model background per bin could fit data well. The backgrounds decrease with increasing $H_T$ rapidly, here we consider two bins 575 GeV$<H_T<$675GeV and $H_T>675$GeV, in which the number of expected background is smaller than 20.  We give the final event number in Fig. \ref{stopcmsn}.

\begin{figure}[!htb]
\begin{center}
\includegraphics[width=0.47\columnwidth]{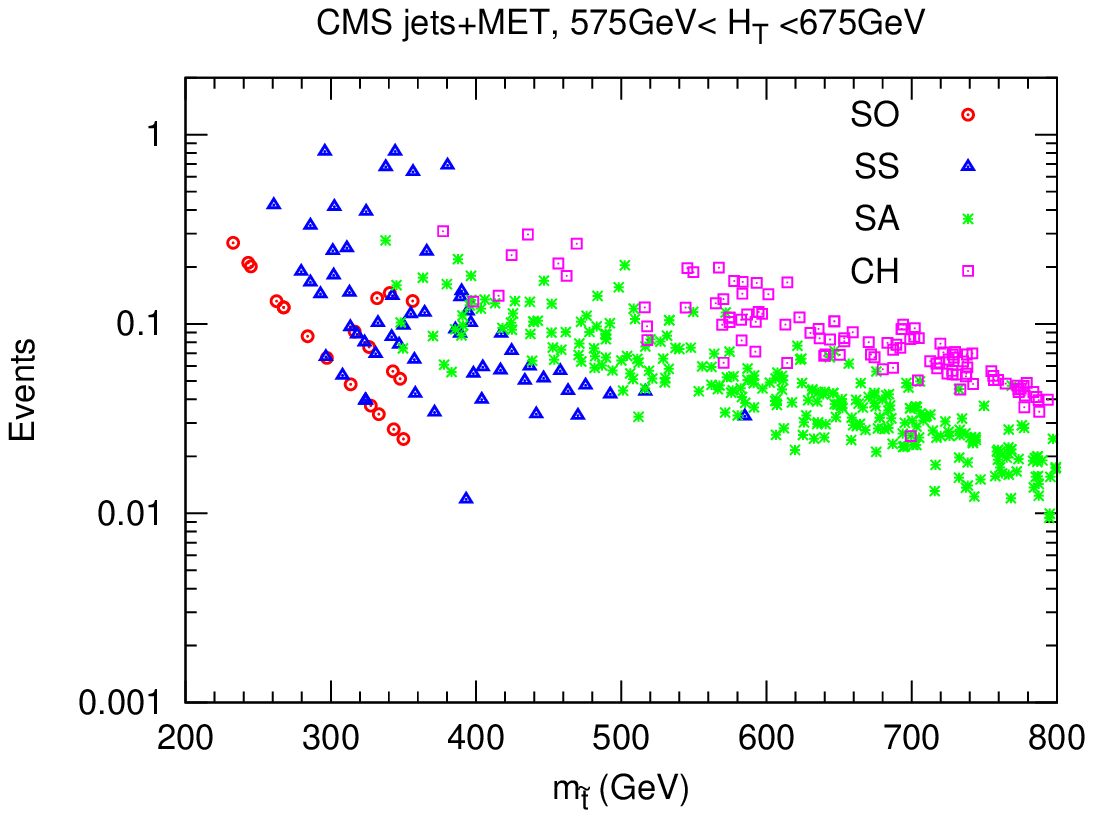}
\includegraphics[width=0.47\columnwidth]{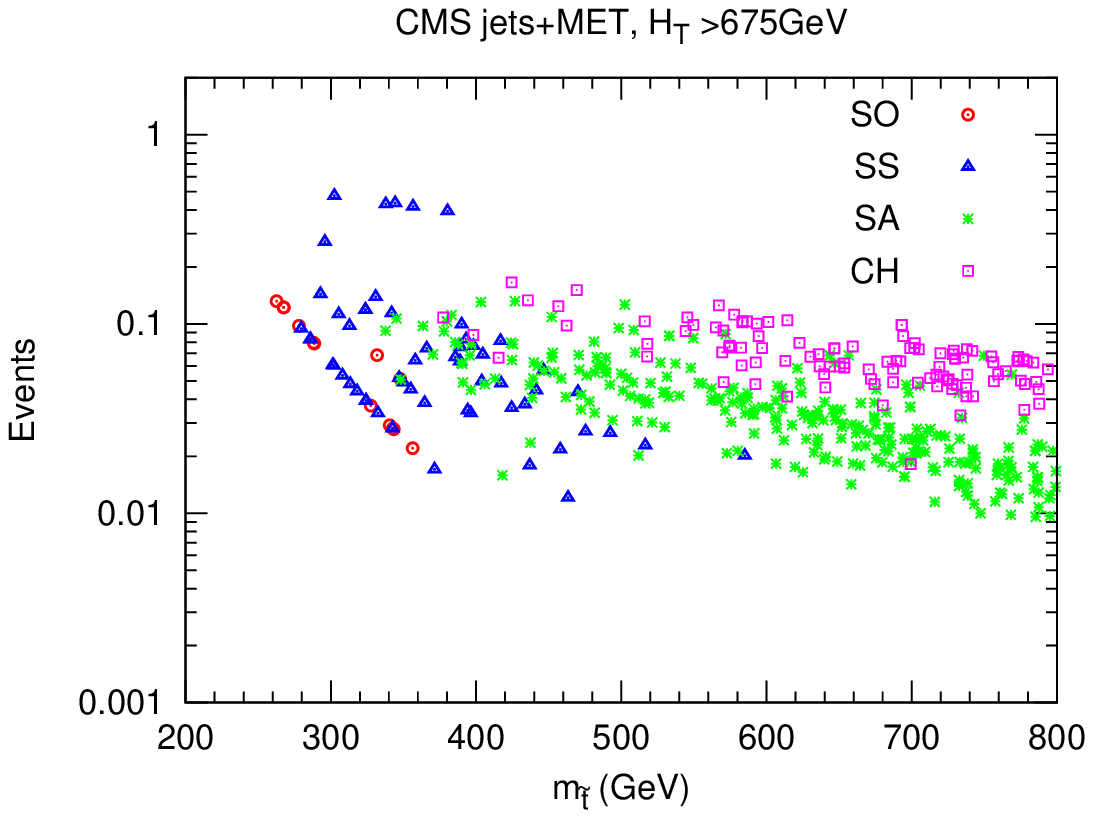}
\caption{The event numbers of signal in the two bins given by the CMS $jets+\sl{E}_T$ searches with 575 GeV$<H_T<$675GeV(left panel) and $H_T>675$GeV(right panel) are shown.
\label{stopcmsn}}
\end{center}
\end{figure}

\item In the Ref. \cite{Aad:2011ib}, ATLAS takes into account the events with jets and  $\sl{E}_T$ with 1.04fb$^{-1}$ of data. Events containing muons with $p_T>$10 GeV, $|\eta|<2.4$ and electrons with $p_T>$20 GeV, $|\eta|<2.47$ are rejected. Reconstructed jets are required to satisfy
\begin{equation}
p_T^{j_1}>130GeV,\;\; p_T^j>40GeV, \;\;\; |\eta|_j<2.8.\;
\end{equation}
The missing transverse energy $\sl{E}_T$ and the minimal $\Delta \phi(j,\;\sl{E}_T)$ between $\sl{E}_T$ and leading jets are required
\begin{equation}
\sl{E}_T> 130 GeV,\;\; \Delta \phi(j,\;\sl{E}_T)_{min}>0.4  \;.\;
\end{equation}
To probe the SUSY parameter points with different gluino and
squark masses, the collaboration define five signal regions as
\begin{itemize}
\item[A] ``2-jet'' region, where cuts are chosen as $n_{j}\geq 2$,
$m_{eff}>$1000 GeV, $\sl{E}_T/m_{eff}>$0.3; \item[B] ``3-jet''
region, where cuts are chosen as $n_{j}\geq 3$, $m_{eff}>$1000
GeV, $\sl{E}_T/m_{eff}>$0.25; \item[C1] ``4-jet'' region, where
cuts are chosen as $n_{j}\geq 4$, $m_{eff}>$500 GeV,
$\sl{E}_T/m_{eff}>$0.4; \item[C2] ``4-jet'' region, where cuts are
chosen as $n_{j}\geq 4$, $m_{eff}>$1000 GeV,
$\sl{E}_T/m_{eff}>$0.25; \item[D] ``High mass'' region, where cuts
are chosen as $n_{j}\geq 4$, $p_T^{j_{2,3,4}}>80GeV$,
$m_{eff}>$1100 GeV, $\sl{E}_T/m_{eff}>$0.2.
\end{itemize}

\begin{figure}[!htb]
\begin{center}
\includegraphics[width=0.47\columnwidth]{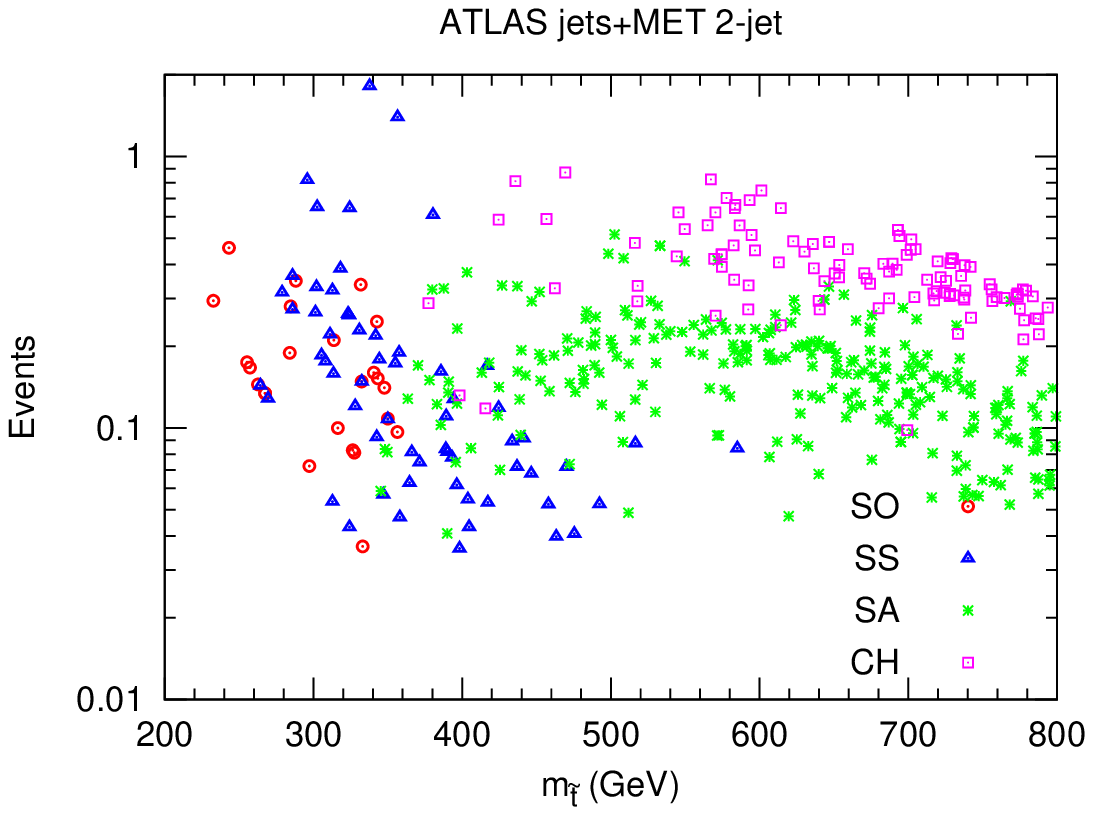}
\includegraphics[width=0.47\columnwidth]{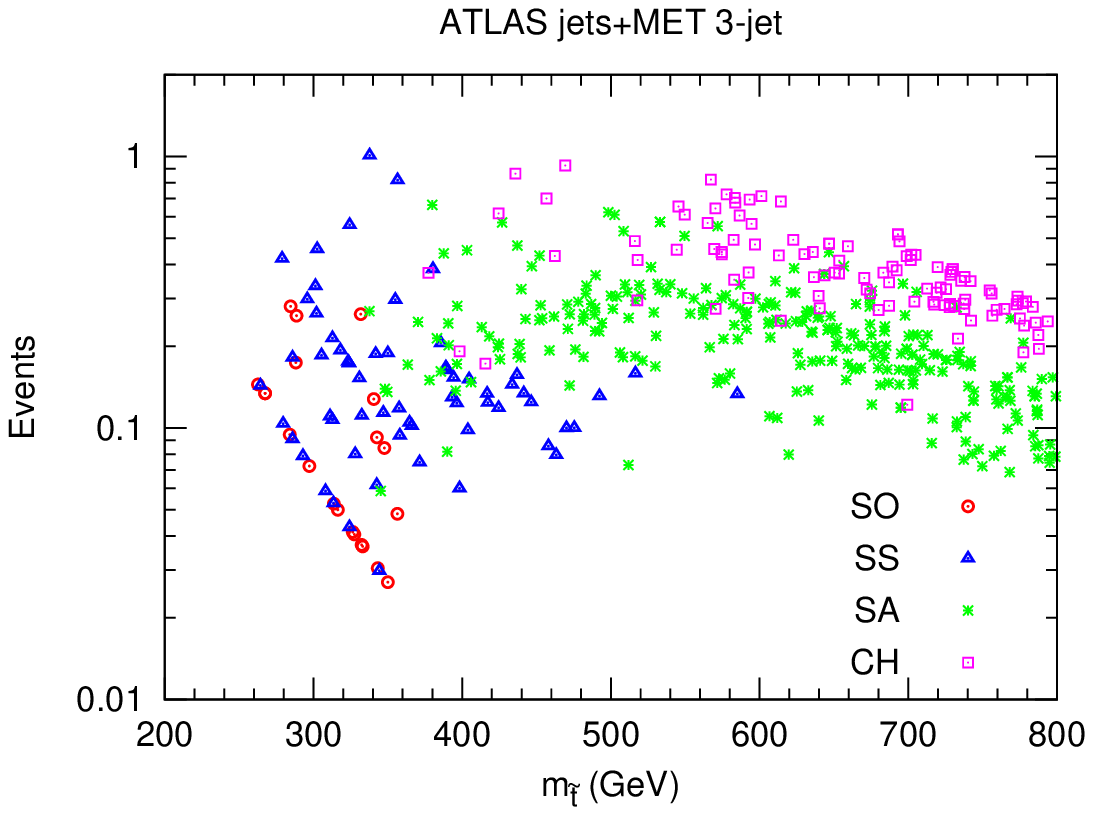}
\includegraphics[width=0.47\columnwidth]{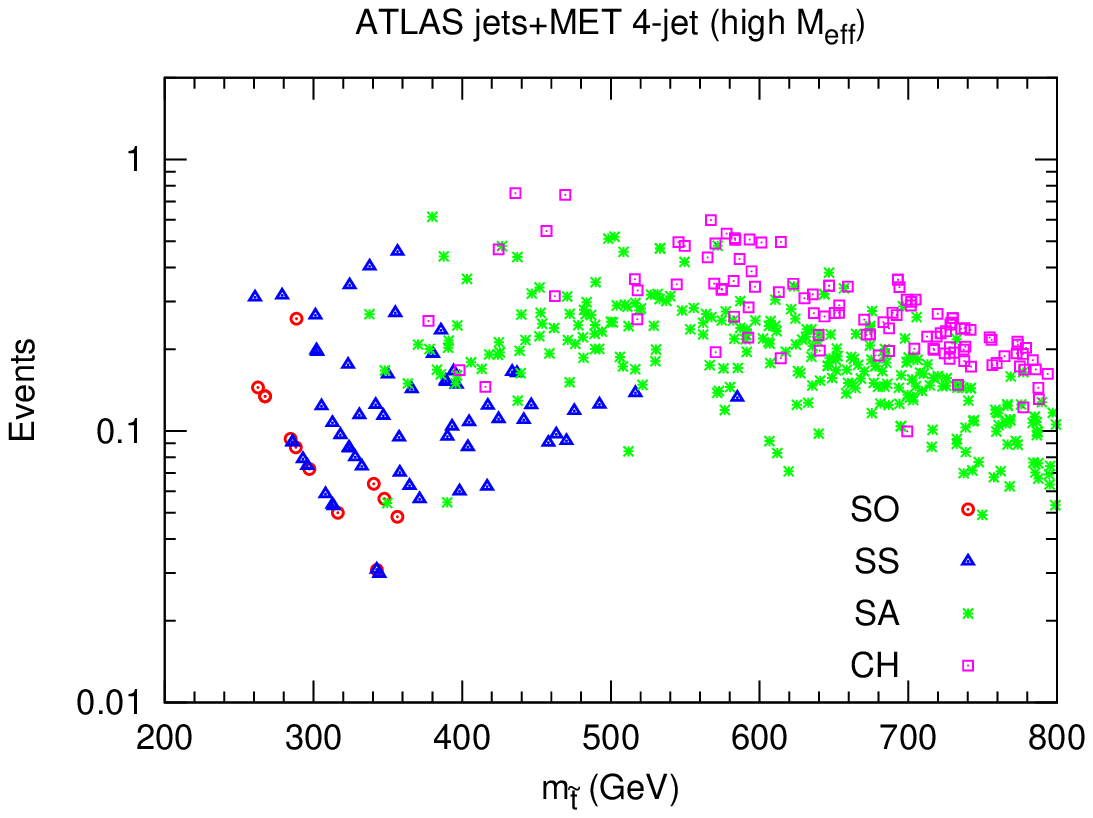}
\includegraphics[width=0.47\columnwidth]{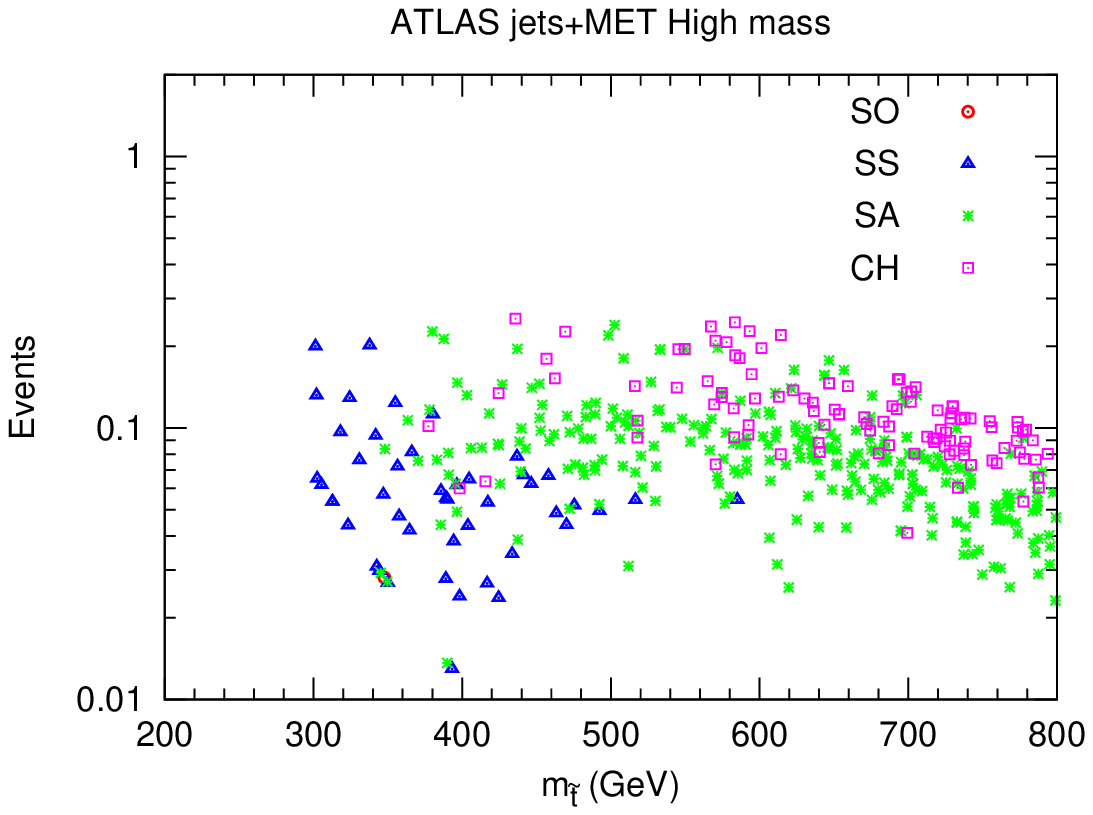}
\caption{The event number of signal after imposing the ATLAS $1bjet+jets+\sl{E}_T$ cuts: "2-jet" region (top left panel), "3-jet" region (top right panel), "4-jet high $m_{eff}$" region (bottom left panel), and "High mass" region (bottom right panel), are shown, respectively.
\label{stopatln}}
\end{center}
\end{figure}

Here $n_j$ is the number of the jets reconstructed in the event, the effective mass is the scalar sum of transverse momentum of jets and the transverse missing momentum $\sl{E}_T$. The signal regions A, B, C and D correspond to different sparticle production channels $\tilde{q}\tilde{q}$,  $\tilde{q}\tilde{g}$, light-$\tilde{g}\tilde{g}$ and heavy-$\tilde{g}\tilde{g}$, respectively. We show the final event number of signal passing all cuts in Fig. \ref{stopatln}. The excluded values of sparticle production cross section are 22fb, 25fb, 429fb, 27fb, and 17fb, respectively.
\end{itemize}

\begin{figure}[!htpb]
\centering
\includegraphics[width=0.32\columnwidth]{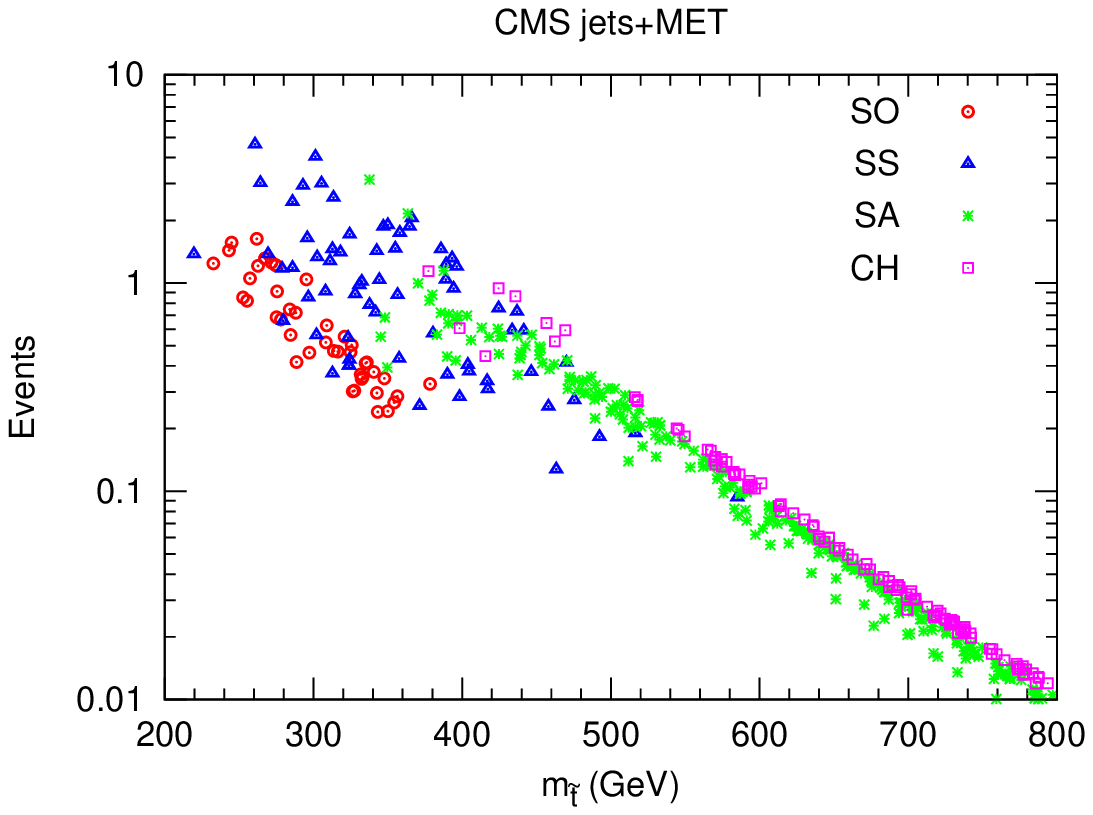}
\includegraphics[width=0.32\columnwidth]{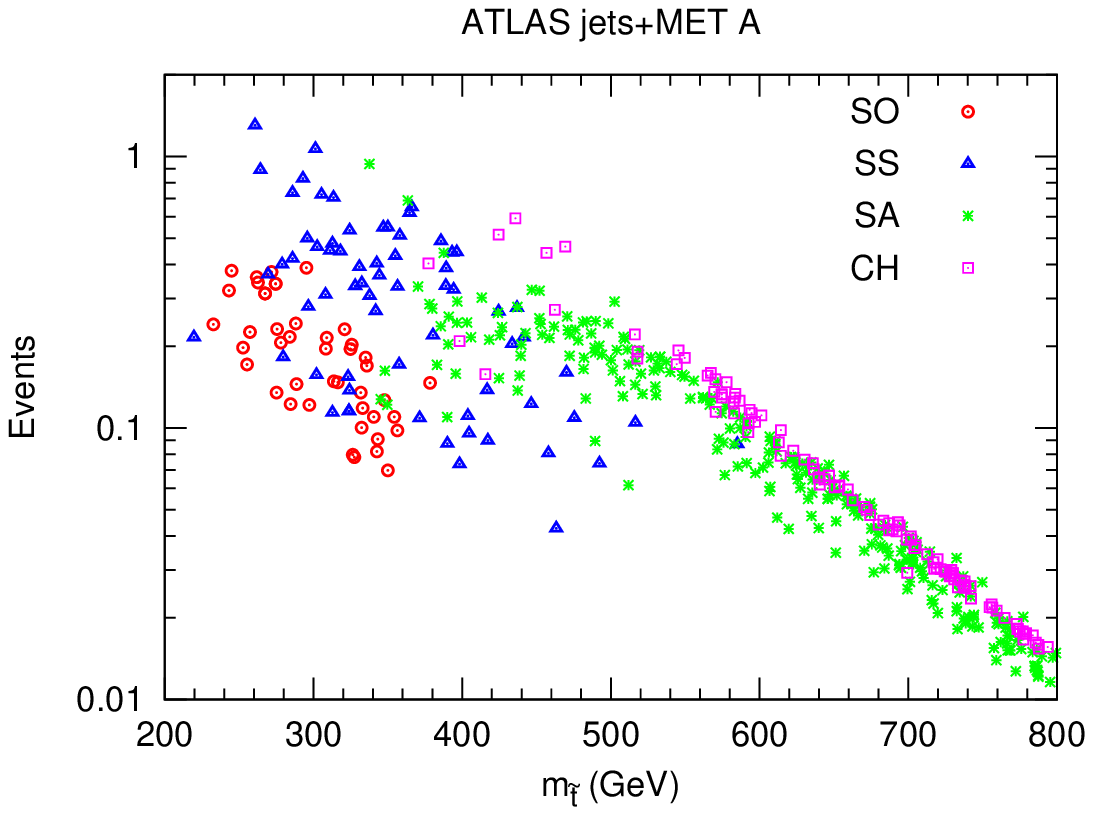}
\includegraphics[width=0.32\columnwidth]{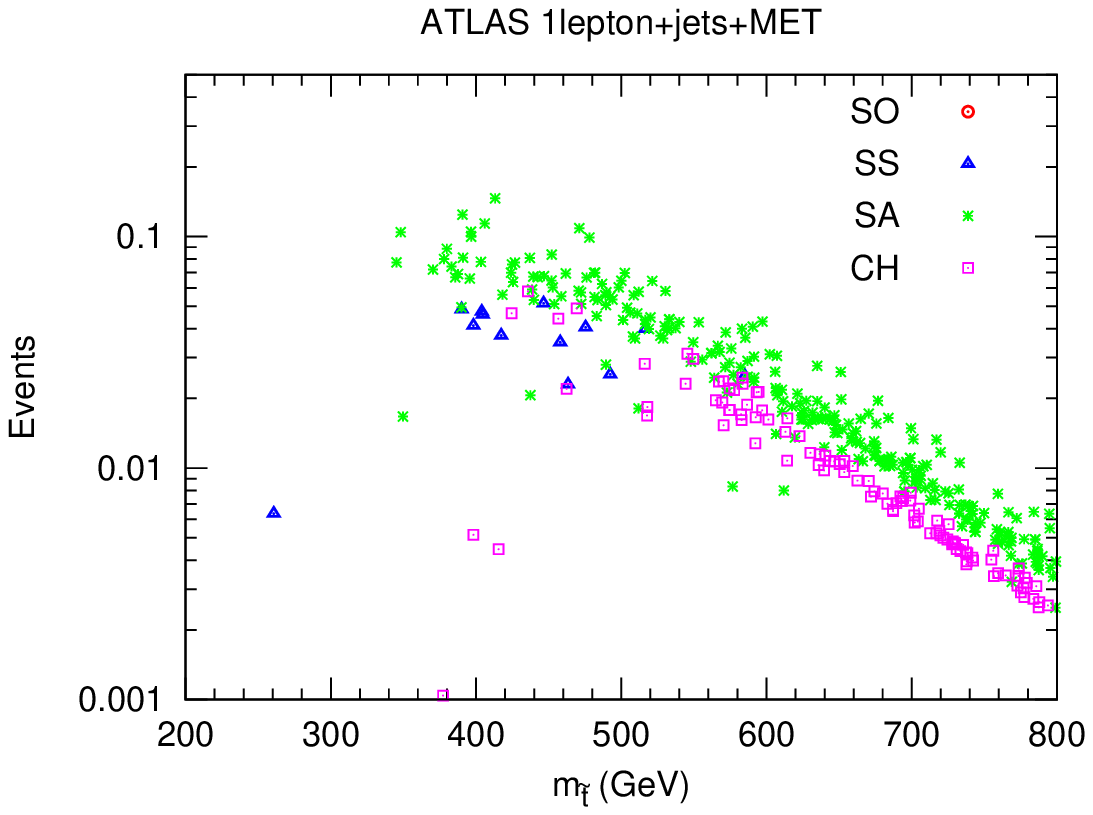}\\
\includegraphics[width=0.32\columnwidth]{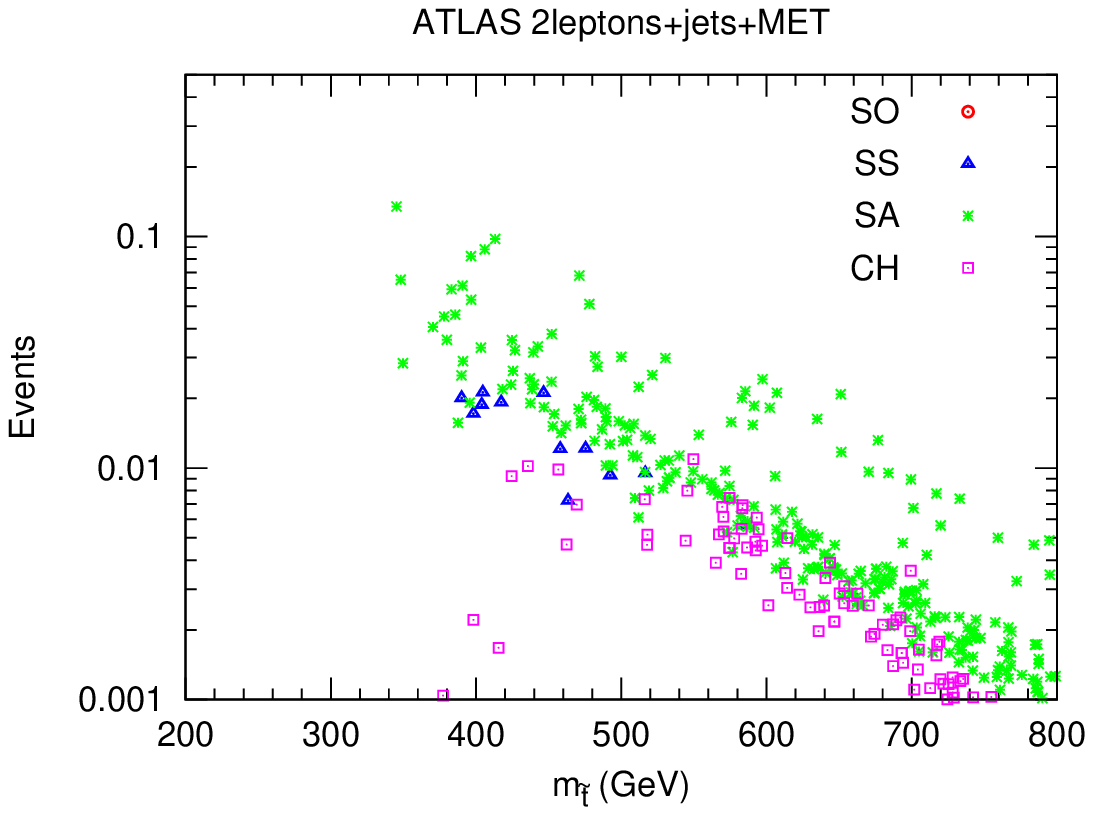}
\includegraphics[width=0.32\columnwidth]{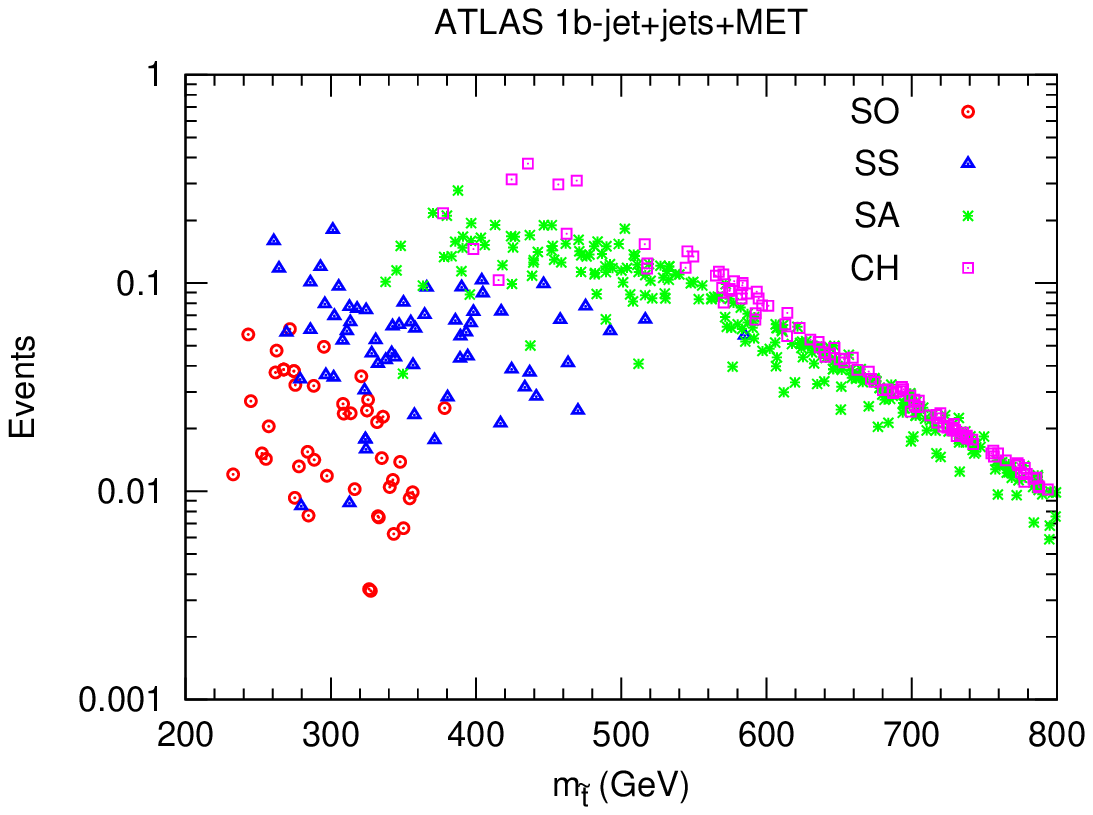}
\includegraphics[width=0.32\columnwidth]{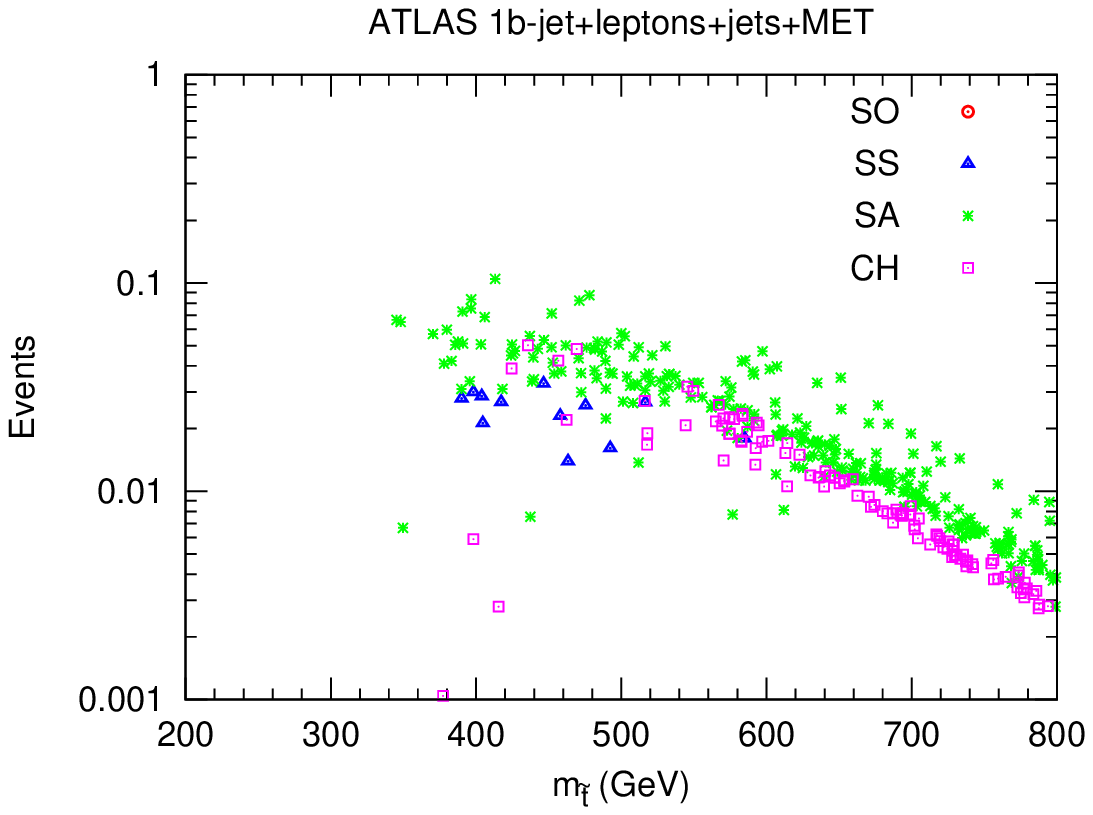}
\caption{The event number of signal passed all cuts from LHC searches with $35 pb^{-1}$ of data (from left to right and top to bottom): (1)CMS $jets+\sl{E}_T$, (2)ATLAS $jets+\sl{E}_T$ (signal region A), (3)ATLAS $1lepton+jets+\sl{E}_T$, (4)ATLAS $2leptons+jets+\sl{E}_T$, (5) ATLAS $1b jet+jets+\sl{E}_T$, and (6) ATLAS $1b jet+leptons+jets+\sl{E}_T$ are shown.
\label{stoplhc}
}
\end{figure}

We also implement the bounds to the stop pair events in term of cuts from the SUSY searches at LHC with 35 pb$^{-1}$ of data. The following channels have been included in our study : (1)CMS' $jets+\sl{E}_T$ channel \cite{Khachatryan:2011tk}, (2)ATLAS' $jets+\sl{E}_T$ channel \cite{daCosta:2011qk}, (3) ATLAS' $1lepton+jets+\sl{E}_T$ channel \cite{daCosta:2011hh}, (4) ATLAS' $2leptons+jets+\sl{E}_T$ channel \cite{Aad:2011xm}, (5) ATLAS' $1b jet+jets+\sl{E}_T$ channel \cite{Aad:2011ks}, and (6) ATLAS' $1b jet+leptons+jets+\sl{E}_T$ channel \cite{Aad:2011ks}. For the channels contained leptons, we do
not focus on the lepton sign and flavor, and sum all the allowed events together. The bounds to the event numbers of the stop pair production are shown in the Fig. \ref{stoplhc}. For these six channel, the upper limits for new physics event number set by collaborations are 13.4, 45.5, 4.7, 20.7, 10.4 and 4.7 respectively.

When comparing the bounds from 35 pb$^{-1}$, we notice that the updated LHC bounds with 1 fb$^{-1}$ of data could not further constrain on those non-universal models allowed by the bounds from 35 pb$^{-1}$ of data. This is simply due to the fact that these cuts are not optimized to put bounds to the light stop pair production by the experimental collaborations.

\subsection{Benchmark points}
Based on the mass pattern proposed above, we select four bench
mark points (BMPs) to demonstrate their spectra and features at
LHC. We tabulate the stop quark and LSP masses in Table
\ref{crossx} and provide the cross sections of these four bench
mark points with the collision energy 7 TeV and 14 TeV.

\begin{table}[th]
\begin{center}%
\begin{tabular}
[c]{|c|c|c|c|c|c|}\hline
Benchmark points      &   BMP1 & BMP2 & BMP3 & BMP4  \\\hline
$m_{\tilde t_1}$  & 390& 243&264 & 338\\
$m_{\tilde \tau_1}$  & 207& 471&199 & 179\\
$m_{\chi^+_1}$  & 383& 424&356& 337\\
$m_{\chi^0_1}$  & 206& 223& 190 & 176\\ \hline
$\sigma$ at 7 TeV (pb) &  $0.23$  & $3.74$  & $2.33$ &  $0.55$    \\ \hline
$\sigma$ at 14 TeV (pb)&  $2.54$  & $28.42$ & $18.91$ &  $5.46$    \\ \hline
\end{tabular}
\end{center}
\caption{The cross sections of $pp\to {\tilde t_1} {\tilde t_1}^*$ in four bench mark points at LHC for collision energy 7 TeV and 14 TeV are presented, respectively. The masses of light sparticles (in GeV) are also given.}%
\label{crossx}%
\end{table}
We focus on the dominant decay channel of the stop, which are listed below:
\begin{itemize}
\item [BMP1]: the stop dominantly decay via the ${\tilde t} \to t \chi^0$ channel ($Br({\tilde t} \to t \chi^0)=98.1\%$);
\item [BMP2]: the stop dominantly decay via the ${\tilde t} \to c \chi^0$ channel ($Br({\tilde t} \to c \chi^0)=98.7\%$);
\item [BMP3]: the stop dominantly decay via the ${\tilde t} \to b \nu_{\tau} {\tilde \tau} \to b \tau + \sl{E}$ channel ($Br({\tilde t} \to b \tau \sl{E})=96.9\%$);
\item [BMP4]: the stop dominantly decay via the ${\tilde t} \to b $ channel ($Br({\tilde t} \to b W \chi^0) =98.2\%$);
\end{itemize}


\begin{figure}[!htb]
\begin{center}
\includegraphics[width=1.0\columnwidth]{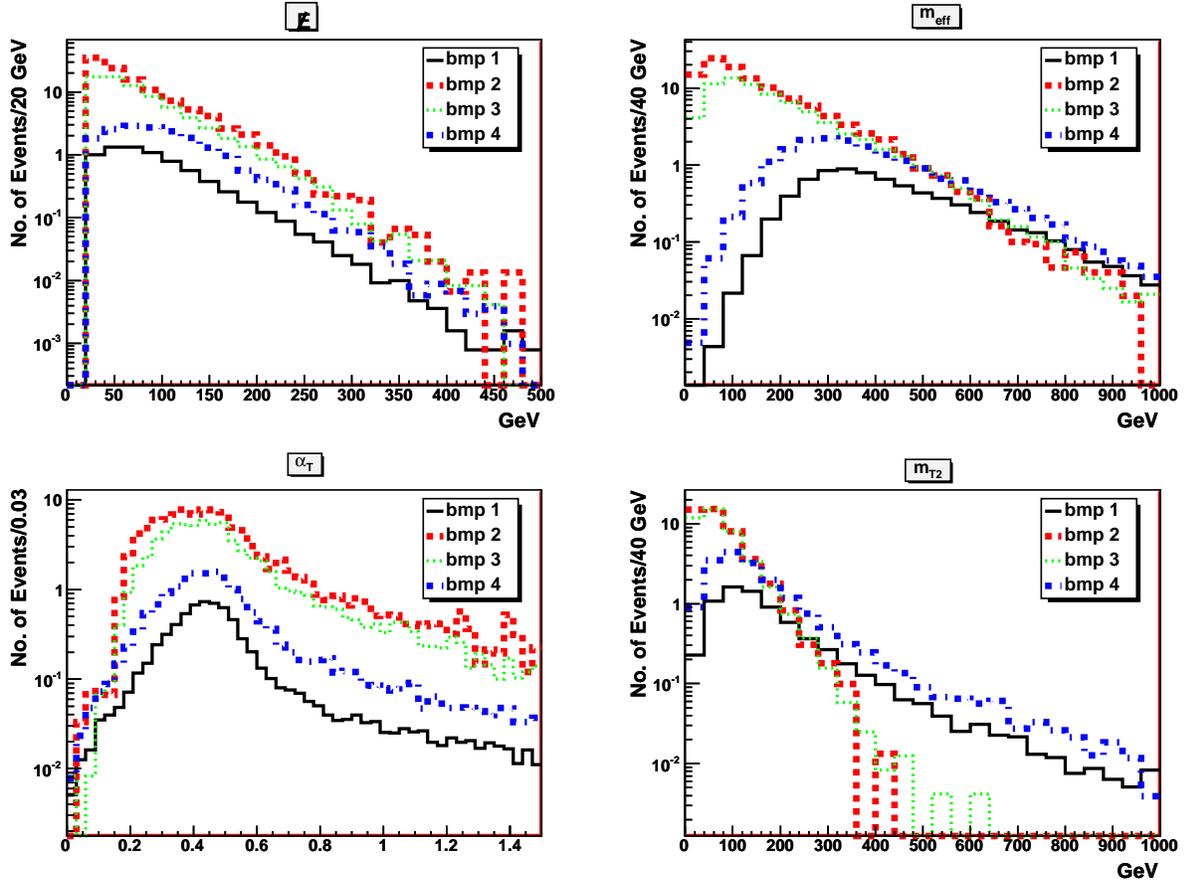}
\caption{The transverse missing momentum, the effective mass of the four bench mark points,  the $\alpha_T$ and the $mT2$ quantity as well are demonstrated.
\label{bmpf1}}
\end{center}
\end{figure}

\begin{figure}[!htb]
\begin{center}
\includegraphics[width=1.0\columnwidth]{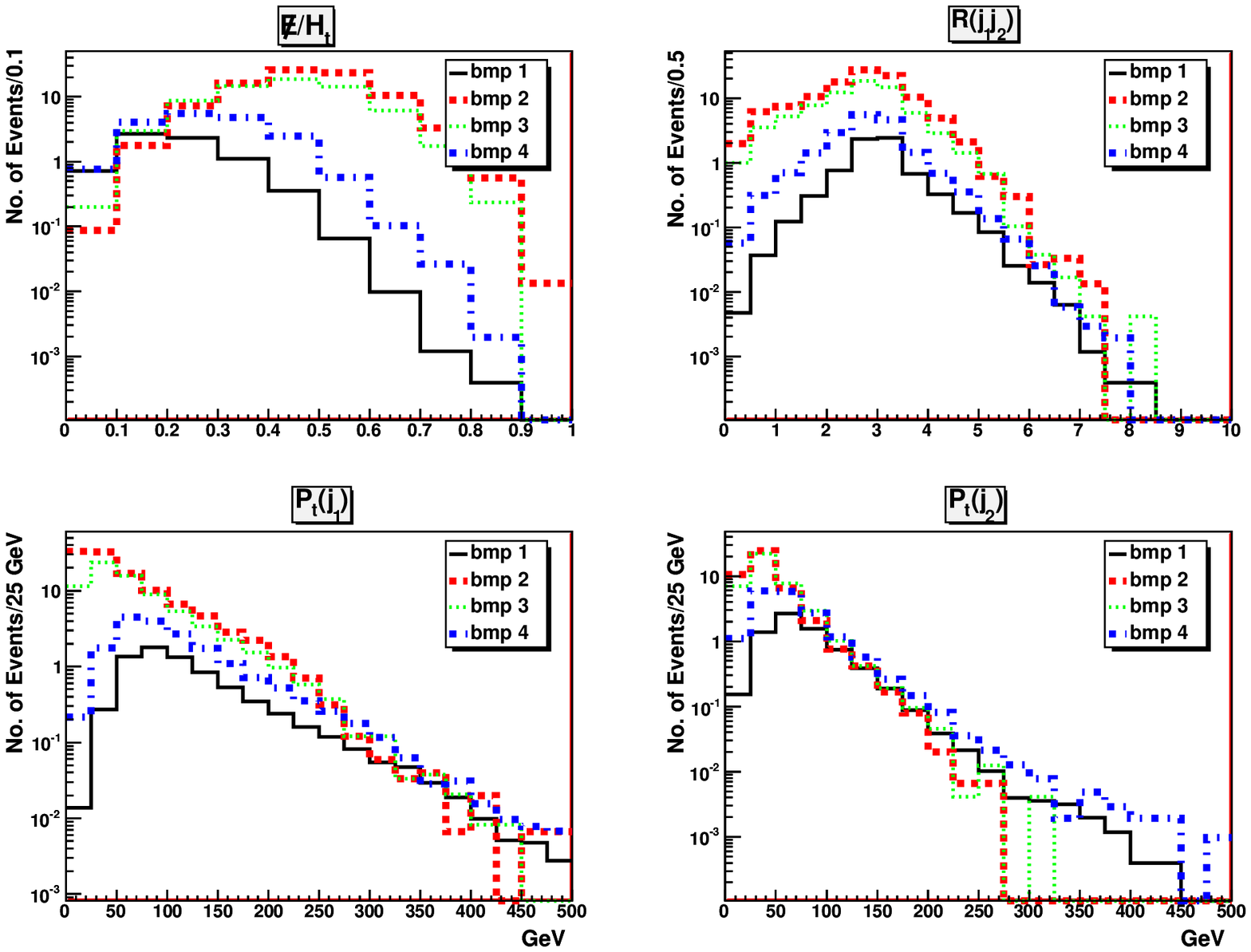}
\caption{ The ratio of missing energy over the $H_T$, the angle
separation of leading two jets $\Delta R(j_1, j_2)$, the
transverse momentum of the leading jet, and the number of jets
distribution in each bench mark point are demonstrated.
\label{bmpf2}}
\end{center}
\end{figure}

Below we show several salient kinematic observables for these $4$
bench mark points in Figs. (\ref{bmpf1}-\ref{bmpf2}). These
observables offer us important clues as how signal can escape the
current LHC SUSY search. The distributions shown in Fig.
(\ref{bmpf1}-\ref{bmpf2}) explicitly show that the current cuts of
ATLAS and CMS which are optimized for the mSUGRA search can not
separate signal events of our bench mark points from the SM
background events. There are several comments in order:
\begin{itemize}
\item The reconstructed transverse missing momentum of these bench mark points is maximal in the region with $\sl{E} < 100 \textrm{GeV}$, as demonstrated in Fig. (\ref{bmpf1}). Therefore the large missing energy cut adopted in experimental collaborations can greatly reduce the signals.
\item The effective mass for the bench mark point 2 and 4 is maximal in the region with $m_{eff}<300 \textrm{GeV}$, while for bench mark point 1 and 4 the $m_{eff}$ can be large enough due to the large stop mass, as shown in Fig. (\ref{bmpf1}).
\item The observables $\alpha_T$ and $mT^2$ are shown in Fig. (\ref{bmpf1}). The observable $\alpha_T$ is supposed to suppress QCD background heavily. Similarly, the razor method \cite{Rogan:2010kb}, which is designed to suppress the huge QCD background at LHC environment and to pick out the signal events from heavy pair-produced particles decay, can not help to distinguish signals from these bench mark points. While the reconstructed $mT^2$ observable \cite{mt2} is not large for all these bench mark points, the survival rate of signal after the cut with $mT^2>300$ GeV can not be larger than $10\%$ for bench mark point 1 and $1\%$ for bench mark point 2, respectively.
\item The ratio of missing energy over $H_t$ is shown in Fig. (\ref{bmpf2}). The experimental cut on this quantity $>0.3$ can affect bench mark point 1 and 4 significantly.
\item The transverse momentum of the leading two jets for the bench mark point is maximal in the region with $P_t < 50$ GeV region for bench mark point 2 and 3, as demonstrated in Fig.  (\ref{bmpf2}). Therefore the cut demanding the leading jet must be larger than $150$ can affect signal significantly. Meanwhile, such a cut can also considerably reduce the signal from bench mark point 1 and 4. Similarly, the cut on the second leading jet can affect all bench mark points badly.
\end{itemize}

Next, we focus on the signature from bench mark point 1 and 2, and
consider the corresponding SM backgrounds. The final states of
these two bench mark points are provided in Table
(\ref{tablebmp1}) and Table (\ref{tablebmp2}), where we have
imposed the acceptance cuts to both of them: 1) $P_t(j)>20$ GeV,
2) $P_t(\ell) > 20 $ GeV, 3) $\sl{E}> 20 $ GeV, 4) $\eta(j) <
2.5$, and 5) $R(\ell,j)>0.4$ and $R(j,j)>0.4$ as well.

\begin{table}[th]
\begin{center}%
\begin{tabular}
[c]{|c|c|c|c|c|c|c|}\hline
                  &  $2j$& $3j$ & $4j$&  $5j$ & $\geq 6j$ \\\hline
$n_\ell=0$ &  $3\%$  & $8\%$   &  $15\%$   & $16\%$   & $19\%$
\\ \hline $n_\ell=1$ &  $3\%$  & $7\%$   &  $7\%$   & $4\%$   &
$2\%$   \\ \hline $n_\ell=2$ &  $0.9\%$  & $0.6\%$   &  $-$   &
$-$   & $-$  \\ \hline
\end{tabular}
\end{center}
\caption{The percentage of lepton and jet multiplicity channels determined by our bench mark point 1 are shown.}%
\label{tablebmp1}%
\end{table}

\begin{table}[th]
\begin{center}%
\begin{tabular}
[c]{|c|c|c|c|c|c|c|}\hline
                  &  $2j$& $3j$ & $4j$&  $5j$ & $\geq 6j$ \\\hline
$n_\ell=0$ &  $20\%$  & $9.5\%$   &  $3.0\%$   & $1\%$   & $-$
\\ \hline
\end{tabular}
\end{center}
\caption{The percentage jet multiplicity channels from the bench mark point 2.}%
\label{tablebmp2}%
\end{table}

Let's first look at the signal from the first bench mark point.
The search strategies for searching ${\tilde t_1} \to t \chi^0 $
can be categorized by the final states in the literature: the full
hadronic channel \cite{Matsumoto:2006ws,Nojiri:2008ir}, the
semi-leptonic channel\cite{Han:2008gy,Alwall:2010jc}, and the
di-leptonic channel. For the hadronic and semi-leptonic modes, some
kinematic observables have been studied to separate the signal and
background, i.e. the missing energy and effective mass. Moreover,
When the tops in the final state are highly boosted, the top
tagger based on the jet substructure analysis can be used to
distinguish signal and background \cite{Plehn:2010st,Plehn:2011tf}. It is also
remarkable that due to the right-handed helicity of the light
stop, the top quarks in the final state should be polarized
\cite{Perelstein:2008zt}. The di-leptonic channel $p p \to {\tilde
t_1} {\tilde t_1}^* \to t {\bar t} \chi^0 \chi^0 \to \ell \ell b
{\bar b} + \sl{E}$ is less studied in literature, the apparent
reason is the small branching fraction. However, consider the
messy background at LHC, two leptons in the final state can help
to suppress background greatly. Furthermore, in order to claim the
signal is from light stop pair decay, di-leptonic channel should
also be observed. So di-leptonic channel is complementary for the
discover of light stop and deserves careful study.

Now, let's look at the signature of the second bench mark point. Search for the signature ${\tilde t_1} \to c \chi^0_1$ at LHC also existed in a vast of literature. If ${\tilde t_1}$ dominantly decays into charm and neutralino, it might be quite challenging to directly detect it even it might be copiously produced, similar decay could occur in the T-parity little Higgs model as demonstrated in \cite{Carena:2006jx}. The charm jet can be very soft which might escape the triggering, meanwhile the reconstructed missing energy can not be large. Therefore such events might not even be recorded. It is known that b-tagging and top-tagging can increase the supersymmetry signal relative the standard model backgrounds. Therefore, c-tagging techniques are also suggested to distinguish signal and QCD background \cite{Kadala:2008uy,Bhattacharyya:2008tw}.

\begin{figure}[!htb]
\begin{center}
\includegraphics[width=0.47\columnwidth]{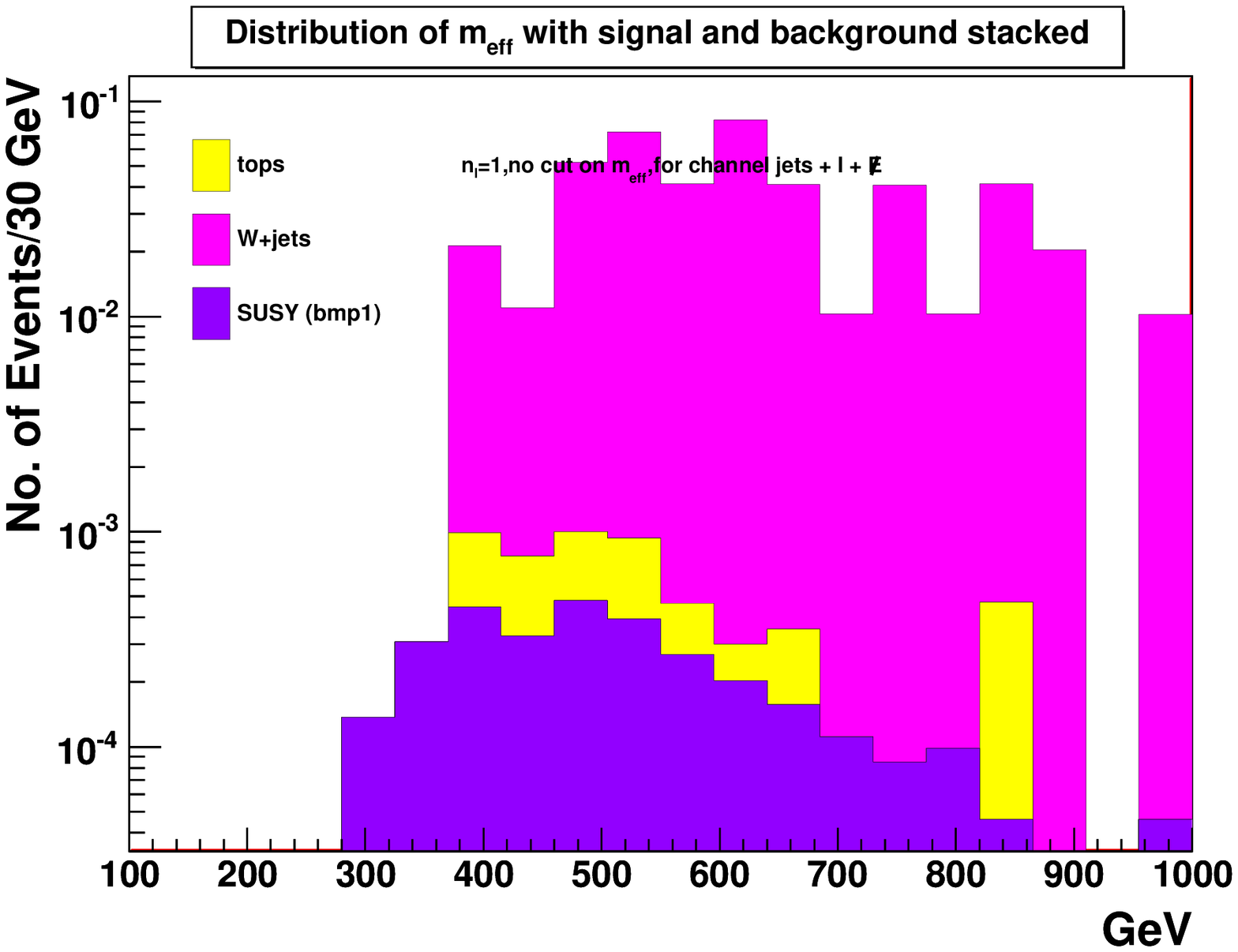}
\includegraphics[width=0.47\columnwidth]{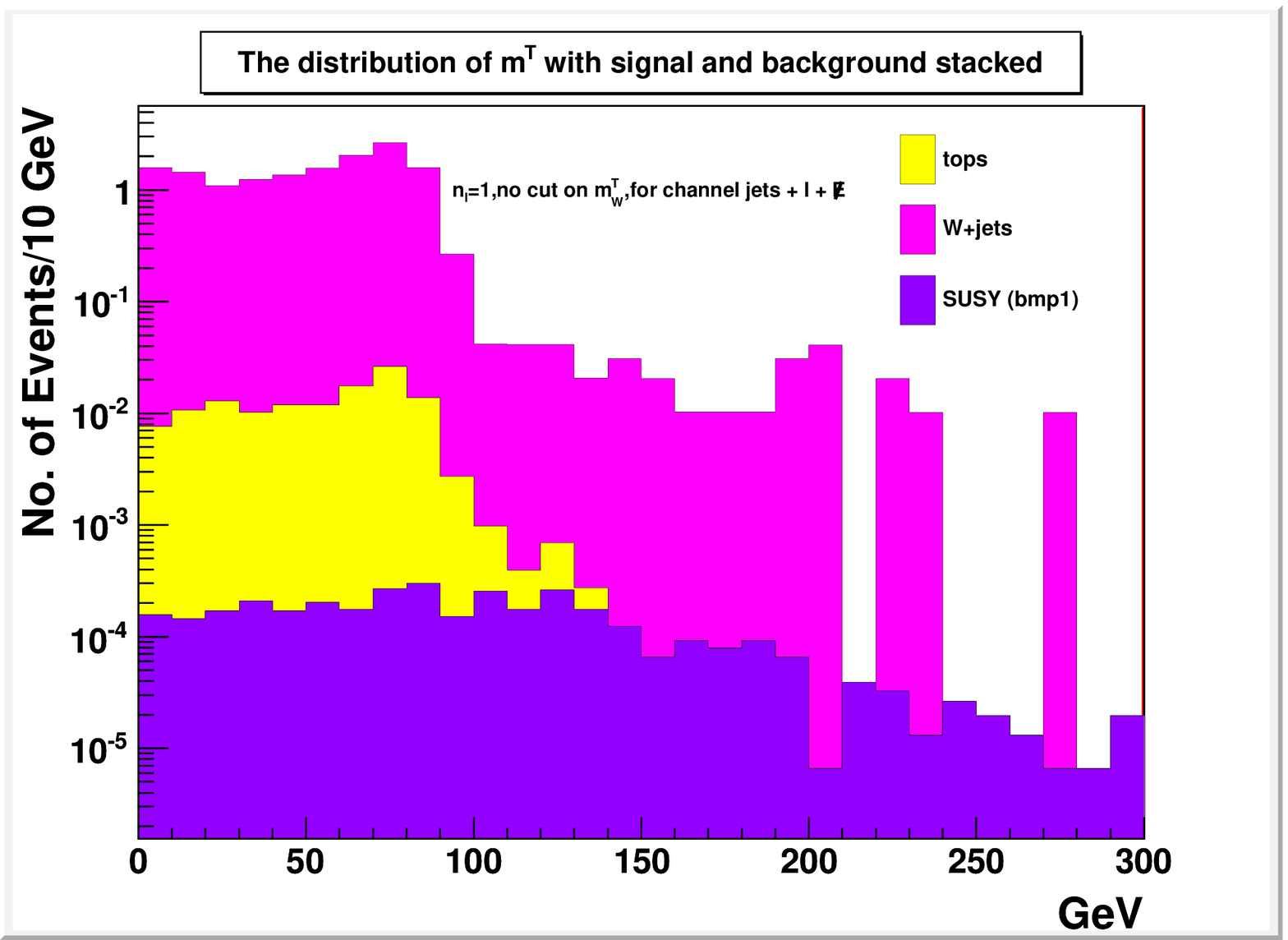}
\caption{The distributions of effective mass $m_{eff}$ and the reconstructed transverse mass $m^T_W$ for the channel $\sl{E} + \ell + jets$ after all other ATLAS cuts for bench mark point 1 in 7 TeV are shown.
\label{bmpf3}}
\end{center}
\end{figure}

\begin{figure}[!htb]
\begin{center}
\includegraphics[width=0.47\columnwidth]{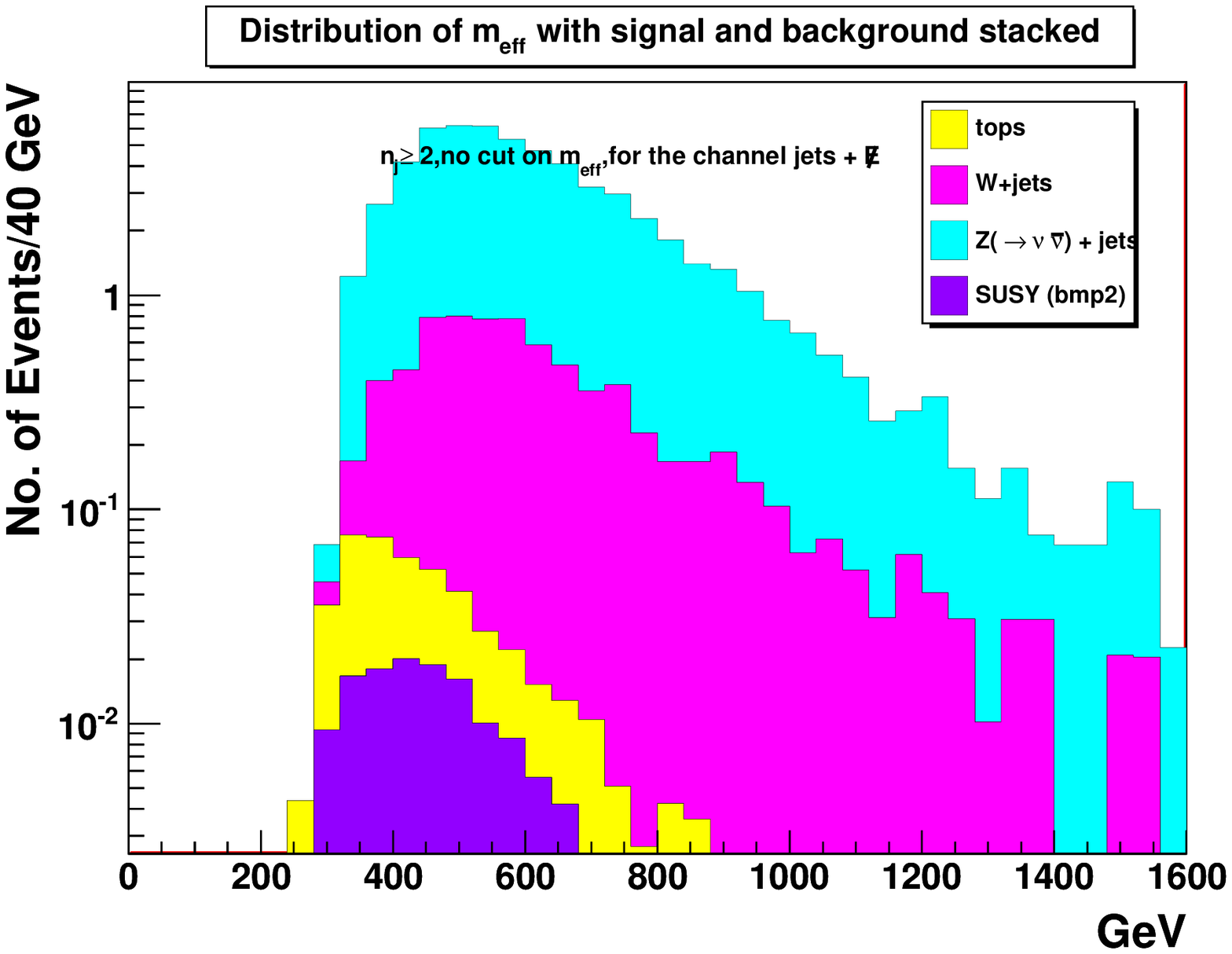}
\includegraphics[width=0.47\columnwidth]{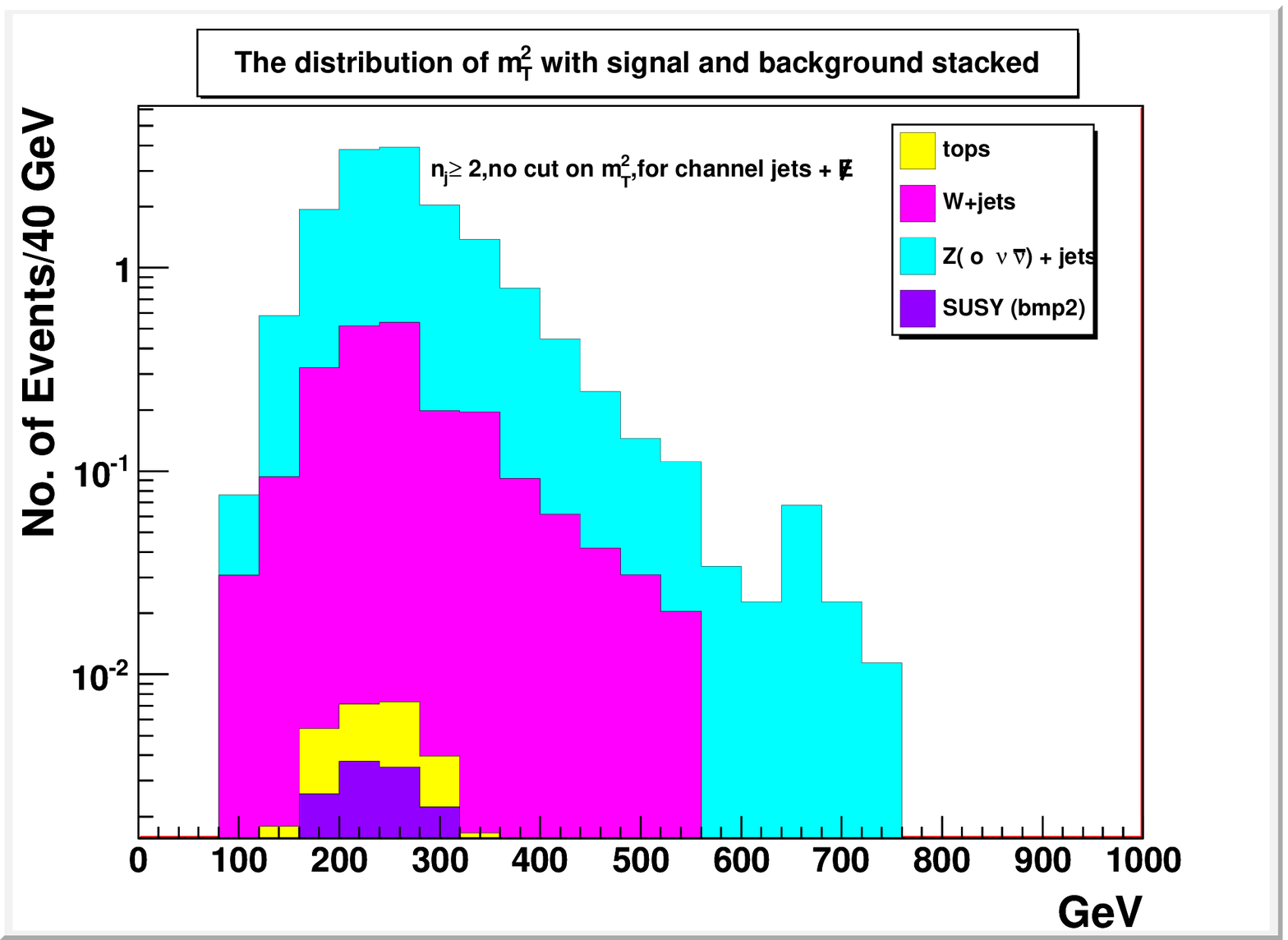}
\caption{The distributions of effective mass $m_{eff}$ and the reconstructed transverse mass $m_T^2$ for the channel $\sl{E} + jets$ after all other CMS cuts for bench mark point 2 in 7 TeV are shown.
\label{bmpf4}}
\end{center}
\end{figure}
In Fig. (\ref{bmpf3}-\ref{bmpf4}) we further show how the current ATLAS cuts and the current CMS cuts affect the observation of our bench mark point 1 and  2, respectively. The QCD background in Fig. (\ref{bmpf4}) can be safely omitted due to the fact that both the $\alpha_T$ and $R_{mis}$ can suppress it significantly. The plots show the distribution of the kinematic observables $m_{eff}$ and $m^T$ with the current ATLAS cuts and  $m_{T}^2$ and $m_{eff}$ with the current CMS cuts. After imposing all cuts, we arrive at the results given in Table \ref{tablebmplhc}. The results indicate that these bench mark points, especially for bench mark point 2, define a  challenge to the current searching strategy. For bench mark point 1, we have studied the full hadronic channel and the di-leptonic channel (mainly suppressed by the branching fraction) and the bound from the semi-leptonic channel is more stringent.

Although the cross section of signal increased by a factor 10 or so from 7 TeV to 14 TeV, the cross section of background also increase with almost the same factor. We arrive at the conclusion that even when LHC can run with the collision energy 14 TeV, it is still challenging to explore these bench mark points.

\begin{table}[th]
\begin{center}%
\begin{tabular}
[c]{|c|c|c|c|c|c|}\hline
                 &   signal  & background & $S/B$ & $S/\sqrt{S+B}$  & Lum. (7 TeV)  \\\hline
BMP1 &  $0.04$  & $4.0$ \cite{daCosta:2011hh} & $1\times 10^{-2}$ &  $0.02$   & $62.5$ fb$^{-1}$  \\ \hline
BMP2 &  $0.01$  & $24.5$ \cite{Khachatryan:2011tk} & $4 \times 10^{-4}$ &  $0.002$   & $6250$ fb$^{-1}$   \\ \hline
\end{tabular}
\end{center}
\caption{Number of events after CMS and ATLAS search cuts with integrated luminosity 35 pb$^{-1}$ are displayed and the required luminosities with the collision energy 7 TeV for the discovery $S/\sqrt{S+B}=5$  of bench mark points 1 and 2 are estimated. }%
\label{tablebmplhc}%
\end{table}
Signature of bench mark point 3 and 4 is $b {\bar b }\tau^+ \tau ^-$, where both b jets and the $\tau$s are typically soft due to the small mass splittings in the cascade decay chains. A recent study on such final states in mSUGRA context can be found at \cite{Bhattacharyya:2011se}. The b-tagging and tau-tagging performance with a soft b jet and a soft tau decreases which may challenge the success to distinguish signal events. Similar to the signature of bench mark point 1 and 2, the current cuts on $m_{eff}$ and $\sl{E}$ suppress the signal badly, therefore to find signature of these two bench mark points at LHC is also difficult. We neglect the detailed analysis of them.

\subsection{Search at the ILC}

The ILC is a future electron-positron collider. At the first stage, it will start with $220$ GeV and run up to the maximal center of mass energy $500$ GeV. At its later stages, it can be upgraded up to $1$ and $3$ TeV. In principle, it is also designed to be able to scan near the threshold region of particle production. Compared with hadronic collider, the ILC enjoys much clean environment as well as high energy resolution capability. For the discovery of SUSY, as one of its advantage, it can reconstruct three dimensional momentum of missing energy.

From the ILC detectors \cite{ilcdetectors,sid,:2010zzd} design report, we take the energy resolution for  jets is assumed as
\bea
\frac{\delta E}{E} = \frac{20\%}{\sqrt{E}} \oplus 1\%\,.
\eea
Meanwhile, since the detectors are supposed to cover the full solid angle, we assume that the coverage of detectors to charged tracks can reach to $20$ for $\eta$ in our fast simulation. Such a detector simulation is realized by modifying the PGS card file.

At the ILC, a pair of light stops can be produced via the
s-channel $\gamma/Z$ exchange.  As pointed out in the design
report \cite{Djouadi:2007ik}, ILC can cover the region of
parameter space with the light stop. Several realistic Monte
Carlo study have been performed in the literature
\cite{Hikasa:1987db,Han:2003qe,Kitano:2002ss,Arhrib:2003rp,Kong:2007uu,Bartl:2009rt}.
Below we focus on the detection of the signal represented by our
bench mark point 2 \cite{Carena:2005gc}.

For the bench mark point 2, the pair production of neutralinos should be possible but can not overcome the background $e^+ e^- \to \nu {\bar \nu}$. Then the light stop should be the first super particle detected at ILC. Here we update the relevant analysis by considering more kinematic observables and using the neural network discriminant to improve signal and background separation. And we find that when compared with the sequential cut method, the neural network discriminant analysis can improve both the ratio of signal over background and the significance remarkably. Our analysis can be extended to the similar study for ${\tilde b} \to b \chi^0$ and we expect the ratio of signal over background and the significance can also be considerably improved. In Figs. (\ref{bmp2ilc-f3}), we show the distribution of key kinematic variables used as the input of the neural network analysis, where we do not impose any a cut except the transverse momentum of jets $P_t(j)>5 \textrm{GeV}$:
\begin{figure}[!htb]
\begin{center}
\includegraphics[width=1.0\columnwidth]{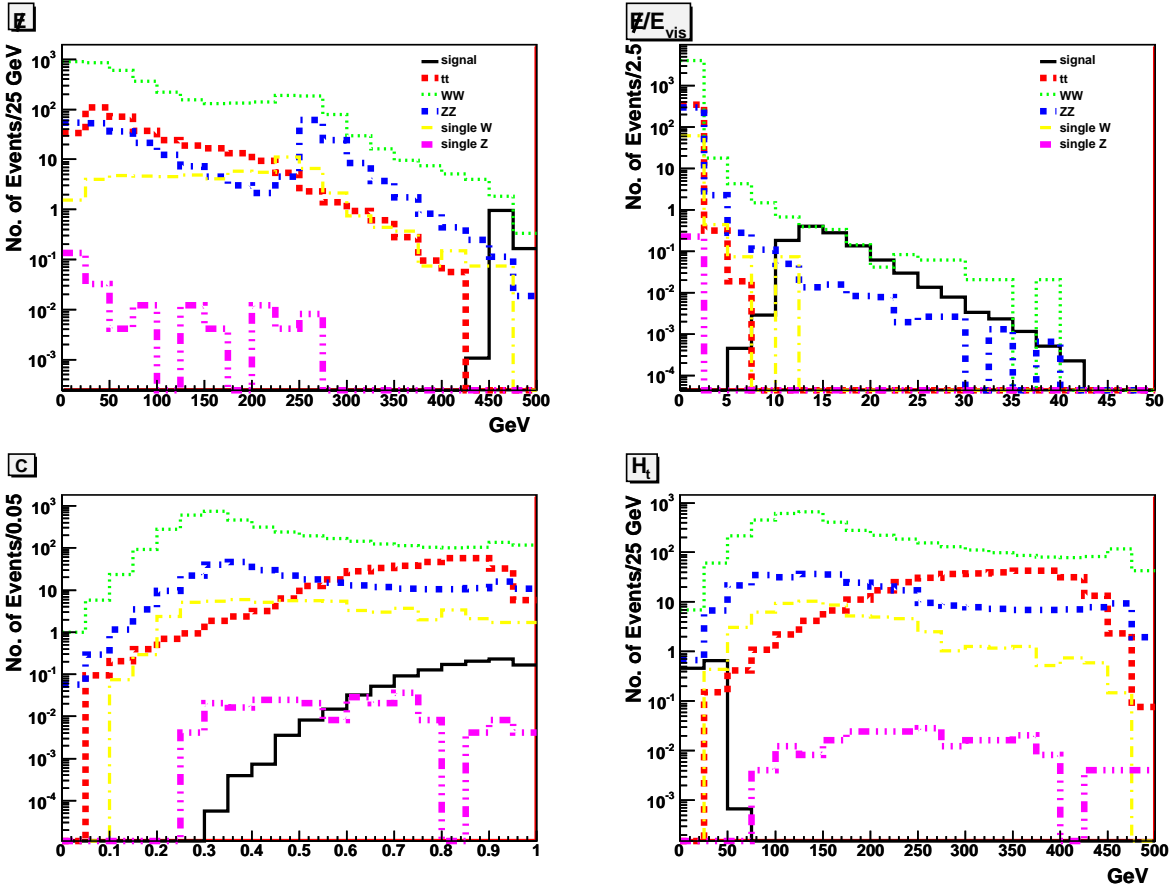}
\caption{ The distributions of the reconstructed missing energy, the invariant mass of leading two jets, the ratio of $\sl{E}/E_{j_1}$, and the invariant mass of missing Lorentz vector as well are demonstrated. The unit of the y-axis is determined by normalizing the integrated luminosity to 1fb$^{-1}$.
\label{bmp2ilc-f3}}
\end{center}
\end{figure}

\begin{itemize}
\item the reconstructed missing energy $\sl{E}$, which is obtained from $\sqrt{s}-E_{j_1} - E_{j_2}$. This quantity can not be reconstructed at LHC but can be reconstructed at ILC. For signal this quantity should be large, as shown in Fig.  (\ref{bmp2ilc-f3}). While for the background $ZZ$ which is occurred in $t$ and $s$ channel, when one of Z decays invisibly, the missing energy can be around $250$ Gev, this explain the bump in the $ZZ$ background.
\item The ratio of $\sl{E}/m_{eff}$, where $m_{eff}$ is defined as the visible energy sum of all objects in the event. Obviously, for signal, this quantity should be large, as shown in Fig. (\ref{bmp2ilc-f3}). From the distribution of this quantity, it becomes quite clear that the dominant background events are from $WW$ and $ZZ$.
\item The centrality $C$. In the signal, the energy tends to deposit in the direction with $\eta =0$ region. We find this quantity is useful.
\item The transverse momentum scalar sum $H_t$ and the jet mass of the two hemisphere jets. If there are more than $2$ jets, we can use the hemisphere algorithm to group jets into two fat jets. For signal, the invariant mass of each jet should be small. While for background, like the highly boosted weak bosons, the invariant mass must be large. We find that these observables are useful to suppress background.
\end{itemize}

We can choose a few pre-selection rules and adopt a few simple cut
method to suppress background while maintain a good acceptance to
signal. At the pre-selection level, we use lepton veto and a cut
on $\sl{E}>300 GeV$. Then the dominant background after
pre-selection is $WW$ pair and $eeZ$ events. We also list the
results in the simple cut method, where we choose: 1) $\sl{E}>425
GeV$, 2) $\sl{E}/(E_{j_1}+ E_{j_2}) > 10$, and 3) $m(j_1,j_2)<60$.
The results are presented at the third line of Table
\ref{tablebmp2ilc}. Due to the correlation among kinematic
observables, it is difficult to find the best set of cuts. To
finish such a task, we utilize the neural network discriminant
analysis to optimize cuts. The results are presented in Table
(\ref{tablebmp2ilc}).

\begin{table}[th]
\begin{center}%
\begin{tabular}
[c]{|c|c|c|c|c|c|c|c|c|}\hline
                 &   signal  & $t{\bar t}$ & $WW$ &  $e {\bar \nu} W$ & $ZZ$ &$eeZ$ &$S/B$ & $S/\sqrt{S+B}$\\\hline
No. of Events after pre-selection &  $11.1$  & $6.2$  & $336.7$ &  $8.9$   & $44.8$  & $-$ & $0.03$  & $0.54$ \\ \hline
No. of Events after a few cuts &  $11.1$  & $-$  & $18.6$ &  $1.0$   & $0.7$  & $-$ & $0.5$  & $1.9$ \\ \hline
No. of Events after NN &  $9.6$  & $-$  & $0.9$ &  $0.7$   & $0.4$  & $-$ & $4.8$  & $2.6$ \\ \hline
\end{tabular}
\end{center}
\caption{The number of events  ( normalized to the integrated luminosity 10 $fb^{-1}$ ) after preselection, after some simple cuts and after the NN discriminant cut are demonstrated. }%
\label{tablebmp2ilc}%
\end{table}

The results of the neural network discriminant are presented at the forth line of Table \ref{tablebmp2ilc}. It is obviously that the NN discriminant analysis can improve both the ratio of signal over background and
the significance. From the results of the neural network analysis, we can estimate the required luminosity is $37$ fb$^{-1}$ for the discovery significance $S/\sqrt{S+B} =5$.

\begin{figure}[!htb]
\begin{center}
\includegraphics[width=0.48\columnwidth]{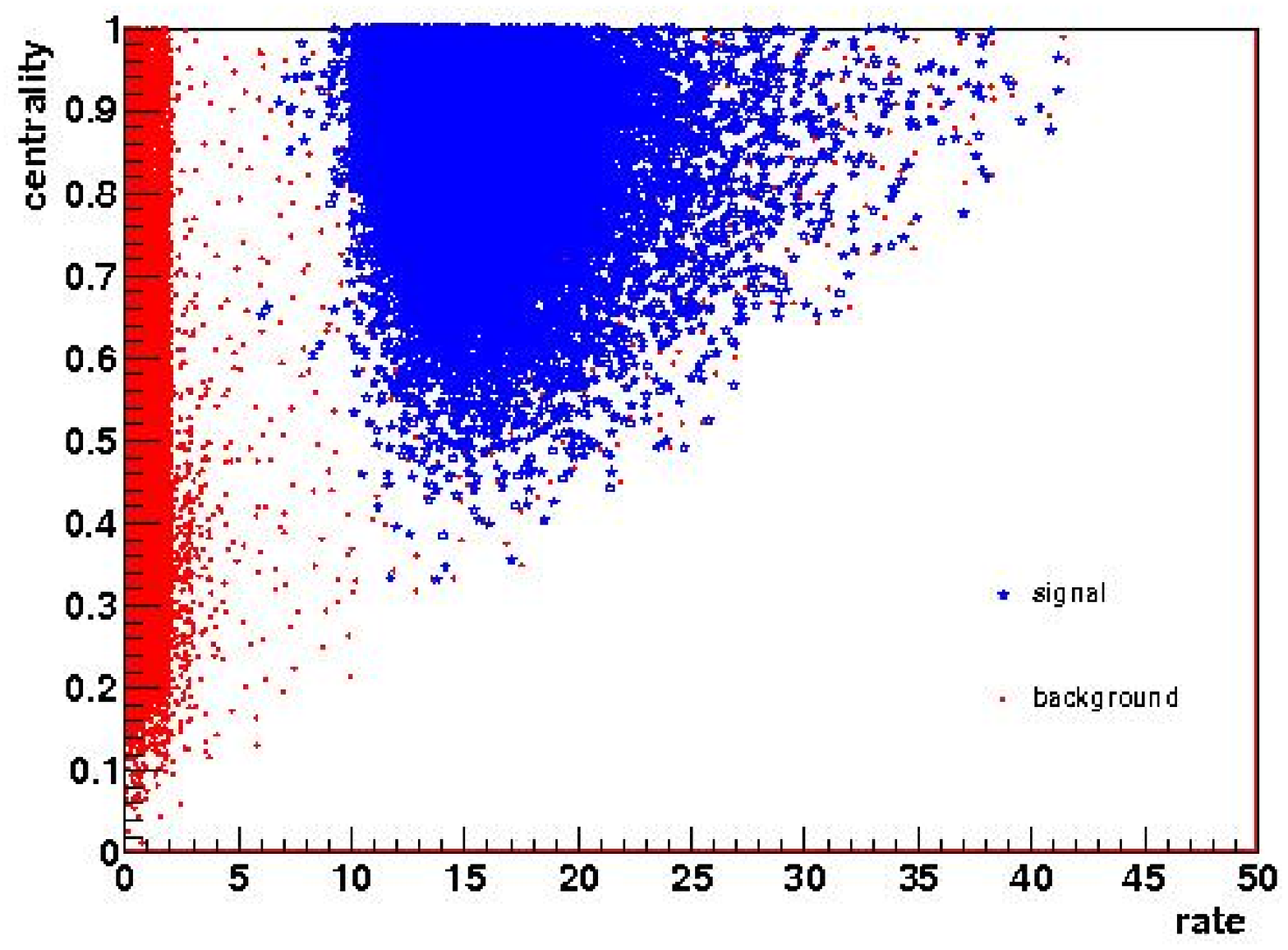}
\includegraphics[width=0.48\columnwidth]{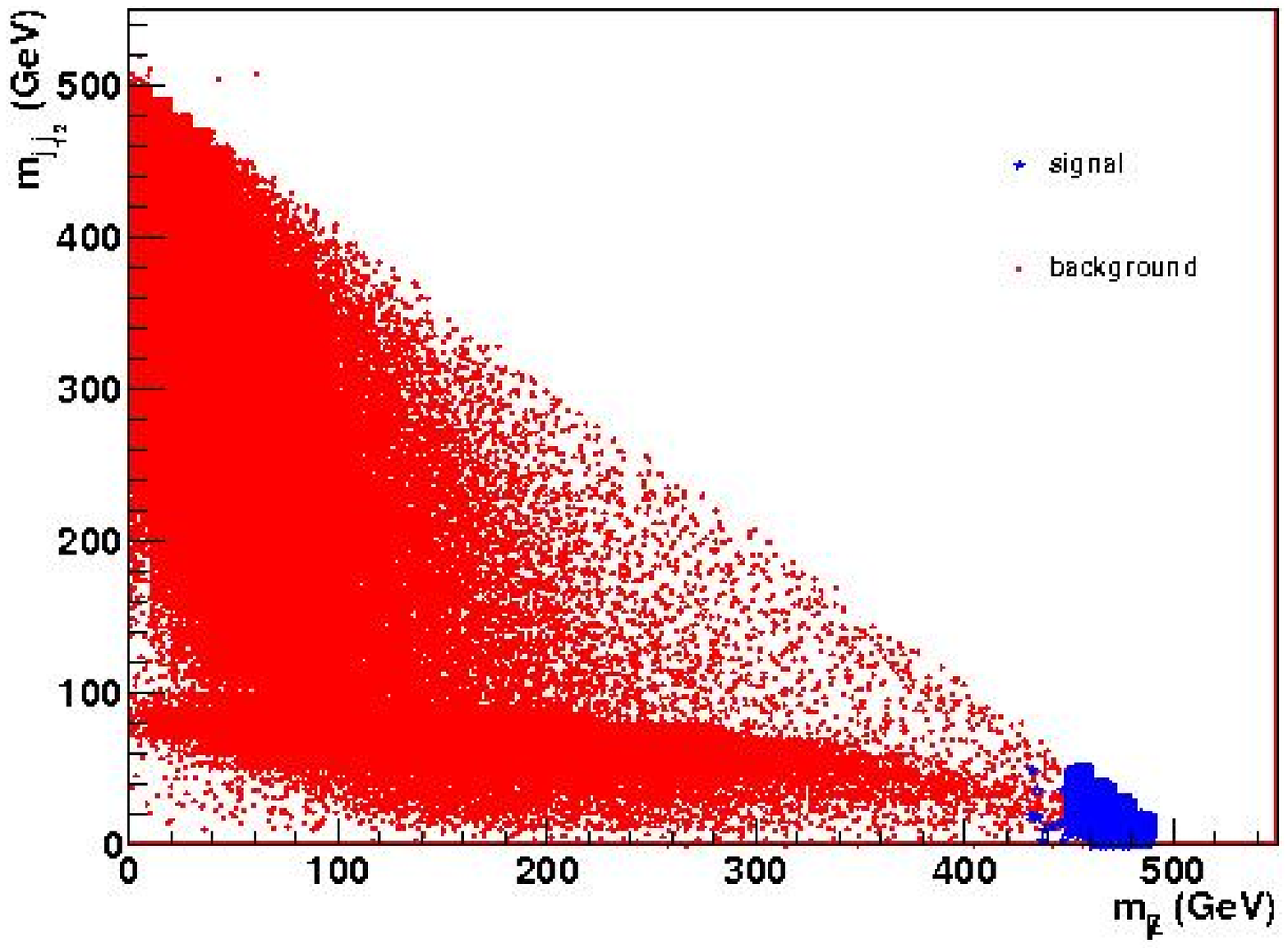}
\caption{Two dimensional scatter plots for signal and background are demonstrated. The blue and red points denote signal and background respectively. In the left panel, the correlations between the rate and centrality for signal and background is shown.
In the right panel the correlation between the invariant mass of the missing 4-momentum and the invariant mass of the two leading jets is shown.
\label{2dob}}
\end{center}
\end{figure}

A simpler version of the neural network discriminant analysis can be demonstrated in Fig. (\ref{2dob}), where two pairs of two almost independent observables are shown. The neural network discriminant analysis basically utilizes such the correlations among observables to distinguish signal and background.

\section{Conclusion and discussions}

We have studied the non-universal SUSY models and explored light
stop pair production at the LHC.
We scan the SUSY parameter space
at the GUT scale and evolve to low energy considering the bounds
by LHC searches with 35 pb$^{-1}$ and 1 fb$^{-1}$ of data  and the
dark matter relic density and direct search bounds. We find that
to give correct relic density the stop usually has small mass
difference with neutralino. Such a scenario easily escape the
current search cuts adopted at both CMS and ATLAS collaborations.

The model we explored in the work is an important scenario
since the colored SUSY particles are the primary goal to search at LHC
and gluino is usually very heavy when evolving from GUT scale to the
low energy scale. Further we have to consider the dark matter relic density
bound, at least considering the upper bound so as not to overclose
the Universe. The dark matter relic density usually leads to degenerate
pattern between the light stop and the LSP neutralino.

In this work we demonstrate that it is difficult to detect the
light stop scenario if only the $pp\to {\tilde t_1} {\tilde t_1}^* $ process
is considered. There have been some studies  to improve the ratio
between signal and background using the associate production, such
as  via associate mono-jet and mono-photon processes \cite{Carena:2008mj,mono-associate}.
It is found by using the associated production,
the large region of the parameter space can be covered by LHC.
Another method proposed to further suppress the background is by
utilizing the two energetic tagged b jets  \cite{Bornhauser:2010mw}.
By studying the $pp\to {\tilde t_1} {\tilde t_1}^* b {\bar b}$ process
\cite{Bornhauser:2010mw}, even the very degenerate
stop neutralino scenario can be explored for LHC at
14 TeV and 500 fb$^{-1}$ integrated luminosity.

Although the associate production channels offer a hope to detect
such a difficult scenario at LHC, the required luminosity seems too large.
Our simulation shows that the ILC is an ideal place to probe these models.
Whether the SUSY is hidden at LHC or SUSY does not exist at low energy
may need more careful and fortitude studies.

\begin{acknowledgments}
The authors thank Qing-Hong Cao, Kai Wang, C. P. Yuan and Shou-Hua Zhu for giving valuable suggestions. This work is supported by the Natural Science Foundation of China under the grant NO. 11075169 and No. 11175251, the 973 project under
grant No. 2010CB833000, and the Chinese Academy of Science
under Grant No. KJCX2-EW-W01.
\end{acknowledgments}

\setcounter{equation}{0}
\renewcommand{\theequation}{\arabic{section}.\arabic{equation}}%

\end{document}